\documentclass[a4paper,fleqn,usenatbib]{mnras}

\usepackage[T1]{fontenc}
\usepackage{ae,aecompl}

\usepackage{graphicx}	
\usepackage{amsmath}	
\usepackage{amssymb}	

\usepackage[export]{adjustbox}

\usepackage[caption=false]{subfig}

\newcommand{\be}{\begin{equation}}
\newcommand{\ee}{\end{equation}}
\newcommand{\bea}{\begin{eqnarray}}
\newcommand{\eea}{\end{eqnarray}}

\def\Sec#1{Section~\ref{sec:#1}}

\def\Fig#1{Fig.~\ref{fig:#1}}
\def\Tab#1{Table~\ref{tab:#1}}

\def\ifm#1{\relax\ifmmode#1\else$\mathsurround=0pt #1$\fi}
\def\kms{\ifmmode\,{\rm km}\,{\rm s}^{-1}\else km$\,$s$^{-1}$\fi}

\def\Msun{\,{\rm M_{\odot}}}

\def\Mbulge{M_{\rm bulge}}

\def\Mstar{M_{\star}}

\def\Mstari{M_{\rm \star,i}}
\def\Mstaripeak{M_{\rm \star,i,peak}}

\def\MBH{M_{\rm BH}}
\def\MBHmed{M_{\rm BH,med}(\Mstar)}
\def\MBHmedipeak{M_{\rm BH,med}(\Mstaripeak)}

\def\Rhalf{R_{1/2,\star}}
\def\Rhalfmed{R_{\rm 1/2,\star,med}(\Mstar)}

\def\tlb{t_{\rm LB}}

\def\ltsima{$\; \buildrel < \over \sim \;$}
\def\simlt{\lower.5ex\hbox{\ltsima}}
\def\gtsima{$\; \buildrel > \over \sim \;$}
\def\simgt{\lower.5ex\hbox{\gtsima}}

\def\r200{r_{200}}
\def\m200{m_{200}}
\def\V200{V_{200}}
\def\M200{M_{200}}
\def\R200{R_{200}}

\def\Vmax{V_{\rm max}}
\def\Vpmax{V_{\rm max,peak}}
\def\Vmaxpeak{V_{\rm max,peak}}

\def\zassstar{z_{\rm assemble, \star}}

\title[Overmassive BHs in EAGLE]{The origin of compact galaxies with anomalously high black hole masses}

\author[C. Barber et al.]{
Christopher Barber,$^{1}$\thanks{Email: \href{mailto:cbar@strw.leidenuniv.nl}{cbar@strw.leidenuniv.nl}}
Joop Schaye,$^{1}$
Richard G. Bower,$^{2}$
Robert A. Crain,$^{3}$\\
\newauthor Matthieu Schaller$^{2}$
and Tom Theuns$^{2}$
\\
$^{1}$Leiden Observatory, Leiden University, PO Box 9513, NL-2300 RA Leiden, The Netherlands\\
$^{2}$Institue for Computational Cosmology, Durham University, South Road, Durham DH1 3LE, UK\\
$^{3}$Astrophysics Research Institute, Liverpool John Moores University, 146 Brownlow Hill, Liverpool L3 5RF, UK
}

\date{Accepted XXX. Received YYY; in original form ZZZ}

\pubyear{2016}

\begin{document}
\label{firstpage}
\pagerange{\pageref{firstpage}--\pageref{lastpage}}
\maketitle

\begin{abstract}

Observations of local galaxies harbouring supermassive black holes (BH) of anomalously high mass, $\MBH$, relative to their stellar mass, $\Mstar$, appear to be at odds with simple models of the co-evolution between galaxies and their central BHs. We study the origin of such outliers in a $\Lambda$ cold dark matter context using the EAGLE cosmological, hydrodynamical simulation. We find 15 `$\MBH(\Mstar)$-outlier' galaxies, defined as having $\MBH$ more than 1.5 dex above the median $\MBH(\Mstar)$ relation in the simulation, $\MBHmed$. All $\MBH(\Mstar)$-outliers are satellite galaxies, typically with $\Mstar \sim 10^{10}\Msun$ and $\MBH \sim 10^{8}\Msun$. They have all become outliers due to a combination of tidal stripping of their outer stellar component acting over several Gyr and early formation times leading to rapid BH growth at high redshift, with the former mechanism being most important for 67 per cent of these outliers. The same mechanisms also cause the $\MBH(\Mstar)$-outlier satellites to be amongst the most compact galaxies in the simulation, making them ideal candidates for ultracompact dwarf galaxy progenitors. The 10 most extreme central galaxies found at $z=0$ (with $\log_{10}(\MBH/\MBHmed) \in [1.2, 1.5]$) grow rapidly in $\MBH$ to lie well above the present-day $\MBH-\Mstar$ relation at early times ($z \simgt 2$), and either continue to evolve parallel to the $z=0$ relation or remain unchanged until the present day, making them `relics' of the high-redshift universe. This high$-z$ formation mechanism may help to explain the origin of observed $\MBH(\Mstar)$-outliers with extended dark matter haloes and undisturbed morphologies.

\end{abstract}


\section{Introduction}
\label{sec:intro}

A growing body of evidence correlating the properties of local ($z\approx 0$) galaxies with their central supermassive black holes (BHs) has been accumulating over the past two decades. Such correlations include relations between the BH mass, $\MBH$, and the host galaxy's bulge luminosity, bulge stellar mass, and stellar velocity dispersion \citep[e.g.,][and references therein]{Kormendy1995,Ferrarese2000, Gebhardt2000,Kormendy2013,McConnell2013}. These correlations are suggestive of co-evolution between the BH and its host galaxy. It is, however, unclear whether there is a direct causal link between them, as in the case of active galactic nucleus (AGN) feedback from the BH acting on the galaxy \citep[e.g.][]{Silk1998, Fabian1999, King2003}, or if they both result from a common physical mechanism such as galaxy-galaxy merging \citep[e.g.][]{Peng2007, Jahnke2011}.

However, every rule has its exceptions. Of the $\sim 80$ galaxies with dynamical $\MBH$ estimates \citep{McConnell2013}, several have been found to host BHs that are approximately an order of magnitude more massive than their bulge luminosities or masses would imply, given the above-mentioned relations. Such BHs have been termed `monsters' \citep{Kormendy2013}, `{\"u}bermassive' \citep{Lasker2013, Ferre-Mateu2015}, `ultramassive' \citep{Fabian2013}, and even `obese' \citep{Agarwal2013}; we refer to galaxies hosting overmassive BHs as $\MBH(\Mstar)$-outliers.  Most notable are the massive elliptical NGC 1277 \citep{VandenBosch2012, Emsellem2013, Fabian2013, Yildirim2015, Scharwachter2016, Walsh2016, Graham2016}, NGC 4486B \citep{Magorrian1998, Gultekin2009, Saglia2016}, and the compact galaxy M60-UCD1 \citep{Seth2014}. All of these observed $\MBH(\Mstar)$-outlier galaxies have been found to lie well above the scatter in the $\MBH$ -- bulge mass ($\Mbulge$) relation, with $\MBH/\Mbulge > 5$ per cent, as opposed to the expected ratio of $\sim 0.3$ per cent \citep{Kormendy2013,McConnell2013}. Other notable, recent examples of $\MBH(\Mstar)$-outliers are NGC 1332 \citep{Humphrey2009, Rusli2011, Barth2016}, NGC 4342 and 4291 \citep{Bogdan2012}, SAGE1C J053634.78-722658.5 \citep[hereafter referred to as S536;][]{vanLoon2015}, MRK 1216 \citep{Yildirim2015}, and NGC 1271 \citep{Walsh2015}, and possibly SDSS J151741.75-004217.6 \citep[hereafter referred to as b19;][]{Lasker2013}.

The presence of such outliers appears to challenge theories of co-evolution between galaxies and their central BHs, and a physical explanation for how they became $\MBH(\Mstar)$-outliers is needed. Two such explanations have been put forward: (1) they formed on the local $\MBH-\Mbulge$ relation, but the galaxies have since been tidally stripped of stars, leaving behind only the galactic core of stars containing a now overmassive BH \citep[e.g.][]{Volonteri2008, Mieske2013, Seth2014}; and (2) they are relics of the early ($z \simgt 2$) Universe when the $\MBH-\Mbulge$ relation may have had a higher normalization \citep[e.g.,][]{Jahnke2009, Caplar2015, Ferre-Mateu2015}. In this latter case the $\MBH(\Mstar)$-outlier galaxies would have formed their stars and BH rapidly at very early times ($z \simgt 2$) and have remained mostly undisturbed until the present day. In this scenario they are expected to have old stellar populations ($\simgt 10$ Gyr old) and to be compact (effective radius less than 2 kpc).

Indeed, both mechanisms may be possible. For example, NGC 1277 has extremely regular isophotes and a flat rotation curve out to 5 times the half light radius, each implying that it is very unlikely to have been tidally stripped \citep{VandenBosch2012}. \citet{Ferre-Mateu2015} also find that NGC 1277, along with six other $\MBH(\Mstar)$-outlier candidates, is compact given its stellar mass and has stellar populations older than 10 Gyr, confirming the likelihood of the `relic' scenario. Indeed, there is some observational evidence that the $\MBH-\Mbulge$ relation was higher at high-$z$, mainly based on the modelling of quasar luminosities and emission lines to measure $\MBH$. However, observational biases and modelling uncertainties make this result highly uncertain \citep[e.g.,][and references therein]{Greene2010}.

On the other hand, the less-massive galaxies NGC 4486B and M60-UCD1 are much more likely to have been tidally stripped of stars, being located a mere 34 and 6.6 projected kpc from much more massive nearby galaxies, M87 and M60, respectively. A stream of globular clusters has been found extending between NGC 4342 and the massive elliptical galaxy NGC 4365, suggestive of severe tidal interactions \citep{Blom2014}. Indeed, one of the favoured theories for the formation of ultracompact dwarf (UCD) galaxies is the tidal stripping of massive progenitors, leaving behind galaxy cores that may contain supermassive BHs \citep[e.g.][]{Bekki2003, Pfeffer2014, Pfeffer2016}. Recently, \citet{Mieske2013} computed stellar masses and dynamical mass-to-light (M/L) ratios for 53 UCDs, finding that their high dynamical M/L ratios (relative to their inferred stellar M/L ratios) can be explained by hypothetical central BHs. Thus the tidal stripping of the stellar component of progenitor galaxies is another promising mechanism for creating perhaps lower mass $\MBH(\Mstar)$-outlier galaxies.

It is also possible that the offsets in the $\MBH$ estimates for some of these galaxies are due to modelling uncertainties (e.g. the assumed stellar mass-to-light (M/L) ratio, initial mass function (IMF), or spatial geometry). For example, using new kinematical maps from the Keck I Telescope combined with Jeans Anisotropic Modelling, \citet{Graham2016} recently computed $\MBH = (1.2 \pm 0.3) \times 10^9 \Msun$ for NGC 1277, an order of magnitude lower than originally estimated by \citet{VandenBosch2012} using Schwarzchild modelling of {\it HST} data. Further examples include the fact that the high dynamical (M/L) ratios of UCDs can also be explained by a variable IMF rather than an overmassive BH \citep{Mieske2013} and that the double nucleus of NGC 4486B has put to question the validity of the spherical isotropic dynamical models used to calculate its $\MBH$ \citep{Gultekin2009}. If analyses of other galaxies also suffer from such uncertainties, the very existence of such $\MBH(\Mstar)$-outlier galaxies seems unclear and thus should be compared with theoretical predictions.

A powerful method of testing scenarios for the formation of atypical galaxies is to look in cosmological simulations of galaxy formation and evolution. In recent years, such simulations have provided both the statistics and the resolution required to study populations of galaxies, within which analogues of these atypical galaxies can be sought. In this paper we use the EAGLE hydrodynamical simulations \citep[hereafter S15 and C15, respectively]{Schaye2015, Crain2015} to investigate first whether such $\MBH(\Mstar)$-outliers are predicted to exist in a $\Lambda$ cold dark matter ($\Lambda$CDM) framework, and if so, to evaluate which physical mechanism, or mechanisms, leads to their existence.

We proceed as follows. In \Sec{EAGLE} we outline the EAGLE simulations used in this paper. \Sec{outliers} describes $\MBH(\Mstar)$-outliers found in EAGLE while \Sec{origin} investigates their physical origins. We relate our results to compact galaxies in \Sec{compactness} and conclude in \Sec{conclusions}.

\section{The EAGLE Simulations}
\label{sec:EAGLE}

The EAGLE project is a suite of state-of-the-art cosmological hydrodynamical simulations with the goal of studying galaxy formation and evolution from shortly after the big bang to the present day. We refer the reader to S15 and C15 for a full description of the simulations, and here provide only a brief overview for completeness.

The EAGLE simulations were run using a modified version of the Tree-Particle-Mesh smoothed-particle hydrodynamics (SPH) code {\sc gadget}-3, last described in \citet{Springel2005}, using periodic boxes with varying sizes and resolutions. The modifications to the SPH implementation are collectively known as `Anarchy' \citep[Dalla Vecchia, in prep; see also][appendix A of S15]{Hopkins2013,Schaller2015b} which alleviates issues with unphysical surface tension at contact discontinuities, includes an improved treatment of artificial viscosity, and a time-step limiter to conserve energy during sudden feedback events. In this paper we focus on the largest EAGLE simulation: the reference (100 Mpc)$^3$ model, simulated  with $1504^3$ particles each of dark matter and gas, with particle masses $(9.7$ and $1.8) \times 10^6 \Msun$, respectively (referred to as Ref-L0100N1504 by S15). A $\Lambda$CDM cosmogony consistent with the {\it Planck} 2013 satellite data release was used \citep[$\Omega_{\rm b} = 0.04825$, $\Omega_{\rm m} = 0.307$, $\Omega_{\Lambda}=0.693$, $h = 0.6777$;][]{Planck2014}. The gravitational softening was kept fixed at 2.66 comoving kpc for $z > 2.8$ and at 0.70 proper kpc thereafter.

The subgrid parameters were calibrated to match the observed $z\approx 0$ galaxy stellar mass function (GSMF), galaxy sizes, and the normalization of the median $\MBH-\Mstar$ relation. In doing so, it has been used to make predictions that match other observables remarkably well, including the Tully Fisher relation and specific star formation rates (S15), the evolution of the GSMF and galaxy sizes \citep{Furlong2015_SFR, Furlong2015_compact}, H$_2$ and H{\sc i} properties of galaxies \citep{Lagos2015, Bahe2016}, the column density distribution of intergalactic metals \citep[S15;][]{Rahmati2016} and of H{\sc i} \citep{Rahmati2015}, galaxy rotation curves \citep{Schaller2015a}, and galaxy luminosities and colours \citep{Trayford2015}. A public data base of the properties of EAGLE galaxies is available at http://icc.dur.ac.uk/Eagle/database.php \citep{McAlpine2016}.  

In Sections \ref{sec:subgrid} and \ref{sec:subfind} we describe the subgrid physics and the method of tracking galaxies in the simulations, respectively.

\subsection{Subgrid physics}
\label{sec:subgrid}

Due to the finite resolution of the simulations, many physical processes that operate on scales smaller than can be simulated accurately (termed `subgrid' physics) are modelled using (analytic) prescriptions. In EAGLE, radiative cooling and photoheating are implemented as per the scheme described by \citet{Wiersma2009a}, where the 11 elements that dominate radiative cooling are followed individually in the presence of the cosmic microwave background and a \citet{Haardt2001} evolving, homogeneous, ionizing UV/X-ray background switched on at $z=11.5$. 

Star formation is implemented with the pressure-dependent star formation law of \citet{Schaye2008} which reproduces by construction the observed Kennicut-Schmidt relation. Gas particles are stochastically converted to star particles when their densities are above the metallicity-dependent star formation threshold of \citet{Schaye2004} which accounts for the metallicity dependence of the density and pressure at which the ISM transitions from a warm, neutral to a cold, molecular phase.  A density-dependent temperature floor corresponding to an equation of state, $P_{\rm eos} \propto \rho_{\rm g}^{4/3}$, with $P_{\rm eos}$ and $\rho_{\rm g}$ the gas pressure and density respectively, is also implemented to guarantee that the Jeans mass of the warm interstellar medium (ISM) is resolved (albeit marginally), thus preventing artificial fragmentation in cold, dense gas \citep{Schaye2008}. 

Each newly formed star particle represents a simple stellar population with a \citet{Chabrier2003a} IMF. Stellar particles lose mass over time according to the metallicity-dependent lifetimes of \citet{Portinari1998}. During the life cycle of a stellar particle, elements are individually injected into the ISM to account for mass loss from core collapse supernovae, winds from AGB stars, and winds from massive stars following the scheme described by \citet{Wiersma2009b}; the mass and energy lost via SNIa are also taken into account. Stellar feedback is implemented by stochastically injecting thermal energy into the gas as described by \citet{DallaVecchia2012}. For each feedback event, the amount of energy injected into each gas particle is kept fixed, but the number of gas particles heated depends on the local gas metallicity and density. The former dependency accounts for the unresolved transition from cooling losses via H and He to the more efficient metal-line cooling at higher metallicity, while the latter prevents excessive artificial thermal losses in high density environments which would otherwise have resulted in the formation of overly compact galaxies (C15, S15). These dependencies were calibrated to match the $z\approx 0$ GSMF and galaxy sizes.

Perhaps most relevant for this paper is the treatment of BHs in the simulation. Once a halo that does not already harbour a BH\footnote{This criterion is necessary since halo mass can fluctuate due to interactions with other haloes.} has reached a total mass greater than $10^{10}$ $h^{-1}\Msun$, it is seeded with a BH by converting the bound gas particle with the highest density into a collisionless BH particle. This particle begins with a (subgrid) BH seed mass of $10^5$ $h^{-1}\Msun$ and grows through mergers with other BHs and accretion of low angular momentum gas, a prescription first introduced by \citet{Springel2005a} and later modified by \citet{Booth2009} and \citet{Rosas-Guevara2015}.  The gas accretion rate is the minimum of the Eddington rate and the \citet{Bondi1944} rate for spherically symmetric accretion, modified to account for the angular momentum of infalling gas \citep{Rosas-Guevara2015}. The BH mass growth rate is then 0.9 times the mass accretion rate, accounting for the assumed radiative efficiency of the accretion disc. BHs are merged when their separation is comparable to the gravitational softening length and their relative velocity smaller than the circular velocity at the smoothing length of the more massive BH. This choice of BH merging model does not affect our results since the galaxies hosting two BHs would be completely merged before the BHs merge and we use the mass of all bound BHs in a halo to define $\MBH$. 

Finally, AGN feedback is performed similarly as done by \citet{Booth2009}. At each time step AGN feedback energy is injected into a subgrid reservoir of feedback energy, which is allowed to heat stochastically the gas neighbouring the BH only after the total energy in the reservoir has reached a high enough value to heat some number of its nearest neighbours by a temperature $\Delta T_{\rm AGN}$, a value that affects the simulated properties of the intracluster medium but is less important for the GSMF (S15). The rate at which the reservoir is filled with energy is proportional to the accretion rate of the BH, with a proportionality constant $\epsilon_{\rm r}\epsilon_{\rm f}$, where $\epsilon_{\rm r}=0.1$ is the radiative efficiency of the accretion disc and $\epsilon_{\rm f}=0.15$ accounts for the fraction of the radiated energy that couples to the gas. As outlined by \citet{Booth2009,Booth2010}, averaged over sufficiently long time-scales, BHs regulate their growth by generating large-scale gas outflows that balance inflows. Since this balance takes place on mass scales much larger than $\MBH$, the energy deposition by the BH required for this balance is not directly dependent on $\MBH$; thus the BH will grow until reaching a critical mass for which the energy output required for self-regulation is reached. Because this critical mass is inversely proportional to $\epsilon_{\rm f}$ for Eddington-limited accretion, this constant was calibrated such that BH masses lie on the $\MBH-\Mstar$ and $\MBH-\sigma$ relations at $z=0$. This point is important, as it implies that the normalization of the $\MBH-\Mstar$ relation in EAGLE is not a prediction, but can be calibrated up or down without affecting the rest of the simulation.

\subsection{Subhalo identification and corrections}
\label{sec:subfind}

Dark matter haloes are identified in EAGLE using a Friends-of-Friends (FoF) algorithm with linking length 0.2 times the mean interparticle spacing \citep{Davis1985} . The \textsc{subfind} algorithm \citep{Springel2001, Dolag2009} is then used to identify self-bound substructures within haloes, termed `subhaloes', using all particle types (i.e., dark matter, stars, gas, and BHs), subject to the requirement that a subhalo must contain at least 20 particles in total. Within an FoF group the central subhalo is defined as the one that contains the particle with the minimum gravitational potential, the others are labelled as satellites. We define a `galaxy' as a subhalo with more than one bound stellar particle. The stellar (BH) mass, $\Mstar$ ($\MBH$), of a galaxy is defined as the total mass of all bound stellar (BH) particles. Note that this definition of $\Mstar$ differs from the mass within a 30 kpc aperture used by S15; however, this choice only makes a significant difference for galaxies with $\Mstar > 10^{11}\Msun$ in the simulation, a mass greater than any galaxies important in this work (S15). Using the mass of only the most massive bound BH for $\MBH$ also does not affect our results. 

Since \textsc{subfind} looks for bound structures, occasionally galaxies can spuriously pop in and out of existence when supermassive BHs in the centres of massive galaxies temporarily become their own bound system, stealing a handful of stars or even the entire galaxy nucleus from the true surrounding galaxy. As mentioned in S15, such artefacts can be prevented by merging subhaloes when one is within both 3 proper kpc and the 3D stellar half-mass radius, $\Rhalf$, of the other. This procedure is crucial to this paper, as here we look specifically for objects that have unusually high $\MBH$, and thus the BH may dominate the total mass of the galaxies in which we are interested. Hence, all $\MBH(\Mstar)$-outlier galaxies presented in this paper were followed through time (in ``snipshots'' with time resolution of $\approx 60$ Myr) to ensure they are not spurious. Indeed, without this step, we find $\approx 30$ subhaloes with $\MBH \sim 10^{9} \Msun$ and $\Mstar \sim 10^{7-8} \Msun$, all but one of which were found to be \textsc{subfind} artefacts. 

Another important issue is that occasionally \textsc{subfind} incorrectly assigns the BH of a satellite galaxy to its host galaxy, temporarily setting $\MBH$ of the satellite to zero. To avoid such incorrect assignments, we reassigned BHs via the following procedure. For each BH, we search for subhaloes for which the BH is within both $\Rhalf$ and 3 pkpc. If such subhaloes exist and the BH's host is not one of these subhaloes, we reassign the BH to the most massive one. This procedure is vital for properly tracking satellite galaxies through the $\MBH-\Mstar$ plane over time (as in \Sec{evolution}), and applies to $\approx 1500$ BHs at the $z=0$ snapshot. Note, however, that none of our $\MBH(\Mstar)$-outlier galaxies (defined in \Sec{outliers}) are affected by this correction at $z=0$.

To track the evolution of individual galaxies through cosmic time, we make use of the merger trees discussed briefly by \citet{McAlpine2016}. The trees were constructed using the algorithm described by \citet{Jiang2014}. In short, a merger tree is constructed by first identifying each subhalo's descendant between consecutive snapshots by tracing $N_{\rm link}$ of its most bound particles (with $N_{\rm link} \in [10,100]$, depending on the number of particles in the subhalo). Descendant-progenitor links are then transformed into a merger tree. Main progenitors are defined as those with the highest branch mass, which is the total mass of all progenitors sitting on a branch beyond some redshift. A full description of the merger trees will be presented by \citet{Qu2016}.

\section{Outliers in the $\MBH-\Mstar$ relation}
\label{sec:outliers}

\begin{figure*}
  \centering
      \includegraphics[width=\textwidth]{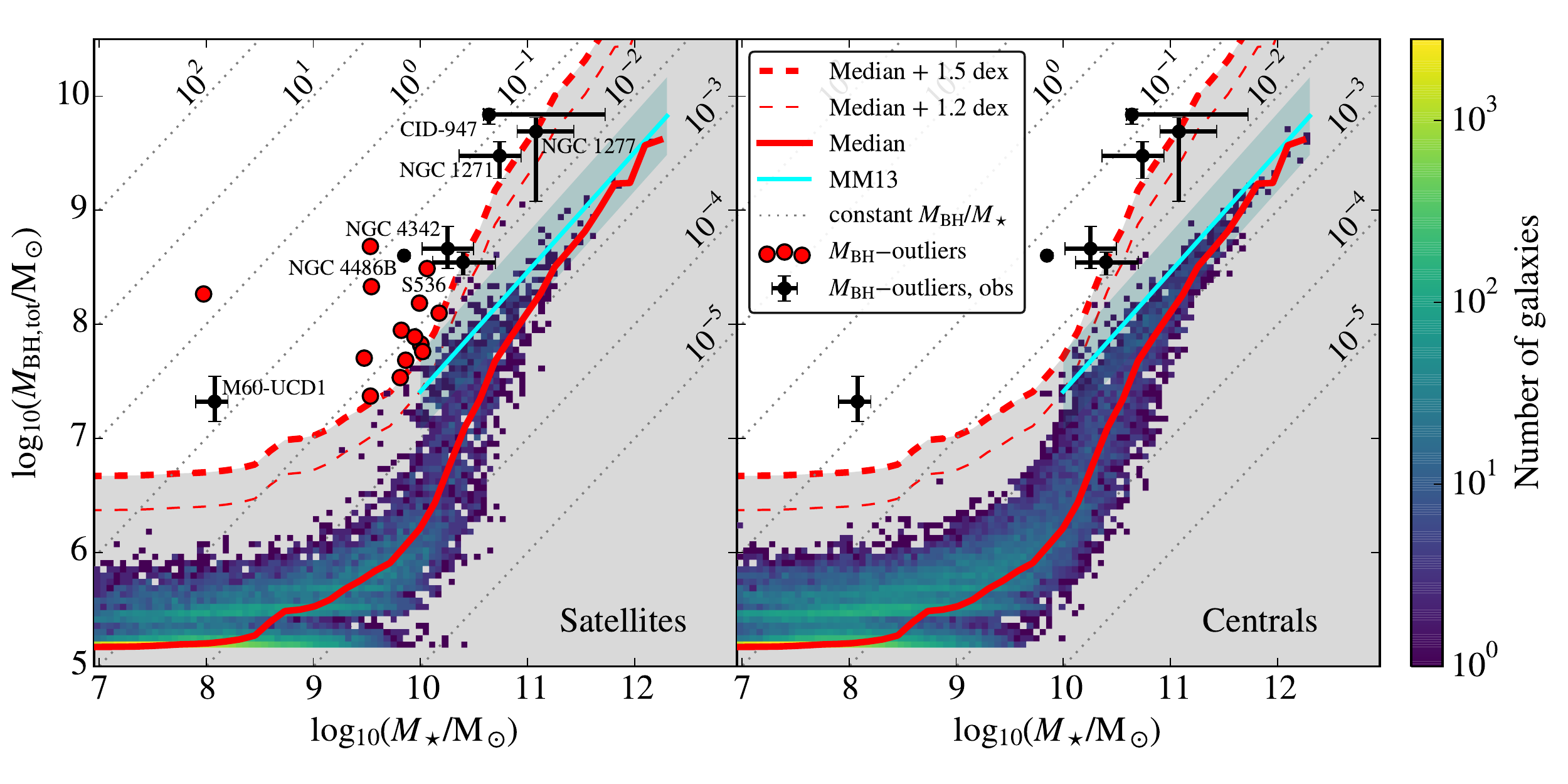}
  \caption{Two-dimensional histogram of BH mass, $\MBH$, as a function of stellar mass, $\Mstar$, in Ref-L0100N1504 for satellite (left) and central (right) galaxies at $z=0$. The median $\MBH$ in bins of $\Mstar$ for all galaxies is drawn as a red solid line in both panels. For reference, the median is redrawn 1.5 and 1.2 dex (see \Sec{others}) higher in  dashed red thick and thin lines, respectively. Lines of constant $\MBH/\Mstar$ are shown as thin grey dotted lines, with corresponding $\MBH/\Mstar$ ratios labelled. We define the 15 galaxies that are at least 1.5 dex above the median to be `$\MBH(\Mstar)$-outlier' galaxies (solid red circles), all of which are satellites, most with $\Mstar \sim 10^{10} \Msun$ and $\MBH \sim 10^{8} \Msun$. The observational $\MBH-\Mbulge$ relation of \citet{McConnell2013} is shown as a cyan line with the intrinsic scatter indicated with a cyan band. Various observed $\MBH(\Mstar)$-outlier galaxies are indicated in black (N.B. for some observed galaxies only $\Mbulge$ data are available; see text for details and references).  }
  \label{fig:MBH_vs_Mstar}
\end{figure*}

\begin{table*}
\caption{Properties of $z=0$ $\MBH(\Mstar)$-outlier galaxies discovered in EAGLE, all of which are satellites. From left-to-right the columns show: OutlierID used in this paper, GalaxyID in the public EAGLE data base, $\Mstar$, $\MBH$, initial stellar mass, peak initial stellar mass, latest redshift at which the galaxy would not be considered a $\MBH(\Mstar)$-outlier, earliest redshift at which $\Mstar(z) > 0.5 \Mstaripeak$, the stellar mass of its host galaxy, the virial mass of its host galaxy, and its separation from the host.}
\begin{center}
\begin{tabular}{rrrrrrrr|rrr}
\hline\hline
OutlierID & GalaxyID & $\Mstar$       & $\MBH$        & $\Mstari $     & $\Mstaripeak $  & $z_{\rm outlier}$ &  $z_{\rm assemble, \star}$ & $M_{\rm \star, host} $ & $M_{\rm 200, host} $ & $D_{\rm host} $ \\ 
          &          & $[10^{9}\Msun]$ & $[10^7\Msun]$ & $[10^{9}\Msun]$ & $[10^{9}\Msun]$ &                 &                        & $[10^9\Msun]$      & $[10^{12}\Msun]$  & [kpc]  \\
\hline

1 & 989925 & 8.85 & 7.75 & 16.13 & 29.12 & 0.3 & 2.0 & 167.89 & 16.1 & 125 \\
2 & 2506301 & 6.48 & 3.41 & 11.63 & 23.61 & 0.1 & 2.0 & 33.96 & 3.2 & 22 \\
3 & 5307530 & 10.05 & 6.56 & 18.36 & 22.23 & 0.9 & 3.0 & 703.05 & 93.9 & 187 \\
4 & 5356576 & 7.29 & 4.84 & 13.38 & 20.25 & 0.9 & 5.0 & 697.19 & 85.6 & 187 \\
5 & 5374884 & 9.80 & 15.34 & 18.06 & 46.24 & 0.1 & 2.5 & 665.39 & 72.7 & 24 \\
6 & 6043240 & 14.94 & 12.53 & 27.44 & 67.81 & 0.3 & 2.2 & 639.12 & 65.5 & 238 \\
7 & 6656922 & 10.09 & 6.72 & 18.25 & 20.50 & 1.3 & 4.0 & 414.12 & 43.6 & 176 \\
8 & 6659446 & 11.54 & 30.82 & 20.81 & 242.76 & 0.1 & 0.4 & 414.12 & 43.6 & 34 \\
9 & 7694028 & 10.48 & 5.77 & 19.15 & 52.76 & 0.1 & 1.5 & 185.33 & 17.5 & 17 \\
10 & 22092348 & 0.09 & 18.44 & 0.17 & 54.88 & 0.6 & 3.5 & 641.28 & 72.3 & 43 \\
11 & 52605772 & 6.64 & 8.86 & 12.02 & 127.74 & 0.1 & 0.6 & 268.16 & 21.0 & 26 \\
12 & 55576881 & 3.41 & 2.35 & 6.23 & 20.03 & 0.1 & 1.5 & 136.74 & 9.2 & 31 \\
13 & 58905530 & 2.99 & 5.05 & 5.48 & 34.31 & 0.3 & 2.2 & 16.42 & 1.6 & 38 \\
14 & 63905307 & 3.49 & 21.39 & 6.40 & 68.46 & 0.1 & 3.5 & 639.12 & 65.5 & 15 \\
15 & 63927059 & 3.41 & 48.19 & 6.25 & 266.94 & 0.1 & 1.3 & 480.14 & 53.4 & 16 \\

\hline
\end{tabular}
\end{center}
\label{tab:outlier_table}
\end{table*}

Since $\MBH(\Mstar)$-outliers are observed, we first ask whether similar outliers exist in EAGLE {\it at all} (and indeed whether we even expect to find them given our resolution and limited volume). \Fig{MBH_vs_Mstar} shows the relation between $\MBH$ and $\Mstar$ for galaxies in the Ref-L0100N1504 EAGLE simulation, separated into satellite (left-hand panel) and central (right-hand panel) galaxies. The median relation for all galaxies in the simulation is shown as a red solid line.  As discussed by S15, the flattening of the simulated data at $\Mstar \ll 10^{10} \Msun$ is due to the BH seed mass of $2 \times 10^{5} \Msun$.  For $\Mstar \simgt 10^{10} \Msun$ the relation steepens due to rapid BH growth, and it flattens slightly above $10^{11} \Msun$.

The observational trend between $\MBH$ and $\Mbulge$ from \citet{McConnell2013} for elliptical galaxies is shown as a cyan solid line. Note that for $\Mstar > 10^{11}\Msun$ most galaxies are elliptical, so here the bulge mass closely approximates $\Mstar$. In this regime, the simulations agree well with the observations considering the $\sim 0.35$ dex intrinsic scatter in the observed trend, shown as a cyan band in \Fig{MBH_vs_Mstar}.

We define $\MBH(\Mstar)$-outliers as those with $\MBH$ at least 1.5 dex above the median $\MBH(\Mstar)$ relation in the simulation [i.e. $\log_{10}(\MBH/\MBHmed) > 1.5$; thick red dashed line in \Fig{MBH_vs_Mstar}]. This criterion was chosen in order to exclude any outliers in the scatter of the low-$\Mstar$ (BH seed mass resolution-dominated) regime, and to select only the most extreme outliers in the simulation\footnote{The scatter around the median $\MBH(\Mstar)$ relation in the simulation peaks at $\Mstar \sim 10^{10}\Msun$, with 68- and 95-percentiles 0.3 and 0.95 dex above the median, respectively. We thus consider our $\MBH(\Mstar)$-outlier definition to be conservative.}. This very simple (and mostly arbitrary) cut leaves us with 15 $\MBH(\Mstar)$-outliers, all of which are satellite galaxies (solid red circles). They have  values of $\MBH = 10^{7-9} \Msun$ and $\Mstar \sim 10^{10} \Msun$, with one interesting case of $\MBH \simeq 2 \times 10^8\Msun$ and $\Mstar \simeq 10^8 \Msun$ which we hereafter refer to as our `most extreme' $\MBH(\Mstar)$-outlier (OutlierID = 10 in \Tab{outlier_table}). Note that a slightly lower choice of $\MBH/\MBHmed$ threshold would add central galaxies to our $\MBH(\Mstar)$-outlier sample; such galaxies are discussed in \Sec{others}. We reiterate here that the absolute value of $\MBH$ is not a prediction of EAGLE $-$ the AGN feedback efficiency was calibrated such that the normalization of the $\MBH-\Mstar$ relation would broadly match observations. This is why we define $\MBH(\Mstar)$-outliers with respect to the simulation only, not to the observations. 

$\MBH$ and $\Mstar$ estimates for various observed galaxies with overmassive BHs are also shown for reference. Values were taken from \citet{Seth2014, Saglia2016, McConnell2013, vanLoon2015, Trakhtenbrot2015, Walsh2015, Walsh2016} for M60-UCD1, NGC 4486B, NGC 4342, S536, CID-947, NGC 1271, and NGC 1277, respectively. Note that for NGC 4486B, 4342, 1271, and CID-947, we plot $\Mbulge$ since total $\Mstar$ is not available, thus $\Mstar$ may be underestimated for these galaxies. For CID-947 (observed at $z\approx3.3$), we plot the expected $z=0$ $\Mstar$ as estimated by \citet{Trakhtenbrot2015} as its upper $\Mstar$ limit. For NGC 1277, we use recent values from \citet{Walsh2016}, but extend the error bars to encompass the new results from \citet{Graham2016}. Overall we see that outliers similar in $\MBH$ and $\Mstar$ to S536, NGC 4486B, and 4342 exist in EAGLE, but we find none similar to those with the largest BH masses: NGC 1277, NGC 1271 or CID-947.  Interestingly, our most extreme $\MBH(\Mstar)$-outlier has similar $\Mstar$ to M60-UCD1, but with an order of magnitude larger $\MBH$.

Given our box size and resolution, it is not surprising that most of our $\MBH(\Mstar)$-outliers have $\Mstar \sim 10^{10} \Msun$ and $\MBH \sim 10^{8} \Msun$. To be an outlier, i.e. to be above the thick red dashed line in \Fig{MBH_vs_Mstar}, the deficit of $\Mstar$ at fixed $\MBH$ increases sharply for $\Mstar < 10^{10} \Msun$ due to the BH seed mass. Thus, in terms of $\Mstar$ deficit, we cannot reliably predict the number of $\MBH(\Mstar)$-outliers for these lower masses. However, in terms of $\MBH$ excess, the lack of $\MBH(\Mstar)$-outliers at $\Mstar<10^{9.5}\Msun$ is significant, since it implies that BHs simply do not grow quickly in such low-mass galaxies without stellar mass increasing even faster. 

At the high-mass end, the number of outliers is affected by the limited box size of the simulation. For $\MBH \in [10^{7.5},10^{8.5}] \Msun$, we find 13 $\MBH(\Mstar)$-outliers out of 389 satellite galaxies. Assuming that the fraction of outliers is the same for all $\MBH$ ($\sim 3$ per cent), for $\MBH \in [10^{8.5},10^{9.5}] \Msun$ we only expect to find one $\MBH(\Mstar)$-outlier since we have have only 33 satellite galaxies in our box with $\MBH$ in this mass range (and indeed, we do find one). Additionally, in our limited volume we have only a handful of objects with $\MBH$ as high as those found in NGC 1271, CID-947, and NGC 1277, so we cannot accurately determine the expected frequency of such objects with this simulation. Indeed, \citet{Saulder2015} found that the frequency of massive, compact, high velocity-dispersion analogues of high-$z$ galaxies (e.g. NGC 1277 and b19) in the local Universe is $\sim 10^{-7}$ galaxies Mpc$^{-3}$ which corresponds to 0.1 galaxies given our simulation volume. Thus, a larger simulation volume would be required to predict the frequency of such galaxies.

Additionally, with higher resolution we may expect to find more $\MBH(\Mstar)$-outliers with $\Mstar \simlt 10^{10} \Msun$ because satellites that lose stellar mass due to tidal stripping by a more massive host are eventually lost by our subhalo finder, perhaps earlier than they would be at a higher resolution. The only EAGLE simulation that has a higher resolution has a particle mass that is 8 times lower than for Ref-L0100N1504 but has a volume of only (25 Mpc)$^3$ (the L0025N0752 simulation in S15) and has only one $\MBH(\Mstar)$-outlier under the above definition, while the (25 Mpc)$^3$ simulation with the same resolution as Ref-L0100N1504 (L0025N0376 in S15) has none. Thus, a robust resolution test is unfortunately not possible. However, once all of the stars have been stripped from a galaxy in the simulation, its BH may still exist as a lone particle, unassociated with any subhalo, prior to merging with the host's BH. We have checked for any such lone BHs and found none. Alternatively, one may expect such BHs to belong to the more massive host prior to merging with its BH. We have searched for BHs with $\MBH > 10^7 \Msun$ at $z=0$ that are {\it not} the most massive BH in their assigned subhalo, finding 99 such BHs with $\MBH$ up to $2\times 10^{9} \Msun$. However, due to the rather ad hoc method of merging BHs in the simulation, it is unclear whether these BHs should be expected to have merged earlier or not.

Each of the $\MBH(\Mstar)$-outlier galaxies was inspected visually in multiple consecutive snipshots to ensure that they are indeed real galaxies and not spurious \textsc{subfind} artefacts missed by the subhalo merging procedure outlined in \Sec{subfind}. Their properties from \textsc{subfind} and those derived in this work can be found in \Tab{outlier_table}. In the next section we describe how these satellite galaxies came to be such strong outliers relative to the $\MBH-\Mstar$ relation.

\section{The origin of outliers from the $\MBH-\Mstar$ relation}
\label{sec:origin}

How did these galaxies become such strong $\MBH(\Mstar)$-outliers? Did they simply form a supermassive BH without forming many stars, or are they the tidally stripped remnants of more massive progenitor galaxies? We discuss their environments at $z=0$ in \Sec{environment}, their evolution through time in \Sec{evolution}, and identify their common origins in Sections \ref{sec:tidalStripping} and \ref{sec:others}.

\subsection{Environment at $z=0$}
\label{sec:environment}

\begin{figure*}
  \centering
  \subfloat{\includegraphics[width=0.3843\textwidth]{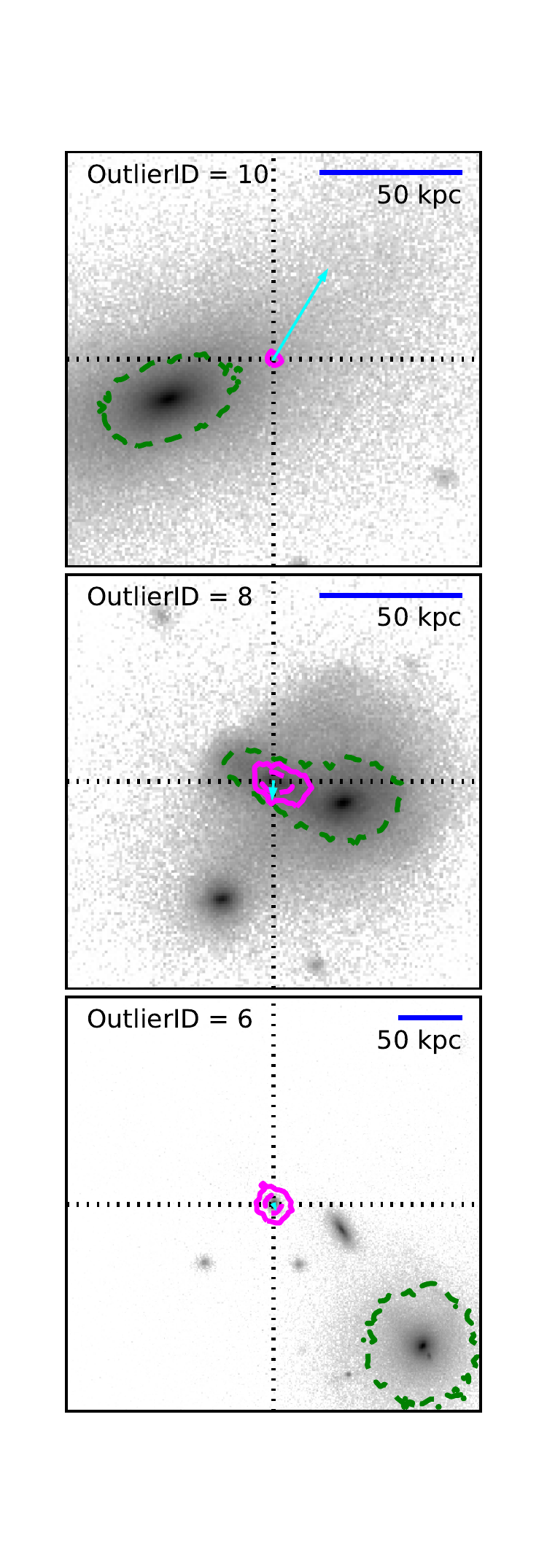} }
  \subfloat{\includegraphics[width=0.52\textwidth]{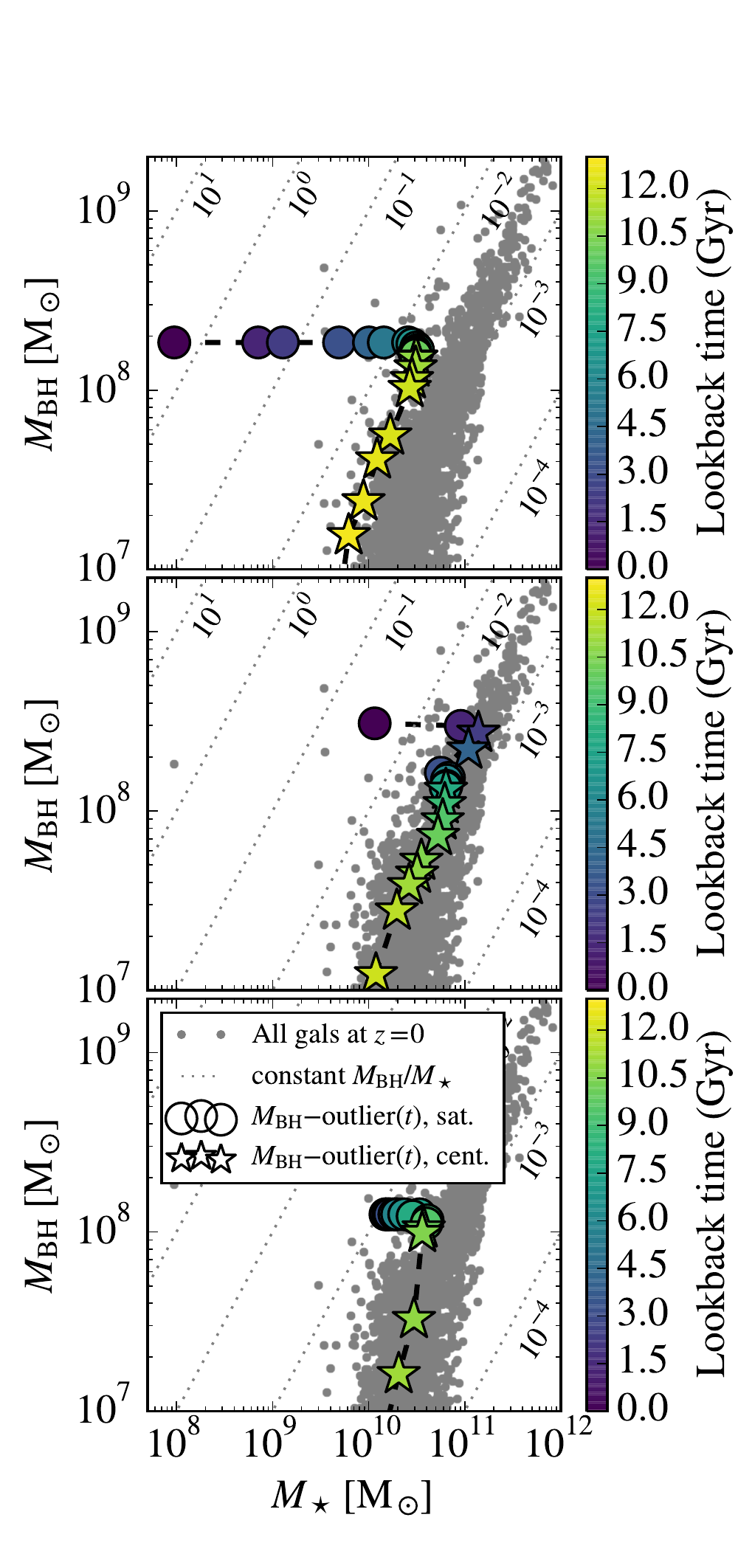} }
  \caption{Environment and evolution of three example $\MBH(\Mstar)$-outlier galaxies (from top-to-bottom: OutlierID 10, 8, and 6 from \Tab{outlier_table}). {\it Left panels:} positions of outliers relative to the nearby stellar particle distribution at $z=0$, centred on their central BHs. The underlying stellar particle distribution is shown as a logarithmic grey-scale surface mass density plot projected $\pm 75$ kpc along the $z$-direction ($\pm 175$ kpc for the bottom panel). In each panel the $\MBH(\Mstar)$-outlier galaxy and its host are outlined in magenta and green contours, respectively, with solid and dashed contours enclosing bins of at least 1 and 100 star particles per pixel, respectively (solid green contours are omitted for clarity). Cyan vectors show the instantaneous velocity in the plotted plane, in units of kpc(50 Myr)$^{-1}$. {\it Right panels:} the evolution of the main progenitors of these three galaxies in the $\MBH-\Mstar$ plane. Symbols are colour-coded by lookback time, $\tlb$, ranging from 12 to 0 Gyr from light yellow to dark blue. Stars and circles show when each $\MBH(\Mstar)$-outlier was a central or satellite, respectively. The underlying distribution for all galaxies at $z=0$ is shown in grey for reference. {\it Top row:} the $\MBH(\Mstar)$-outlier with lowest stellar mass, stripped slowly but substantially over the past $\sim 8$ Gyr. {\it Middle row:} an outlier that has lost $\approx 90$ per cent of its stellar mass within the past 1 Gyr. {\it Bottom row:} an outlier that looks seemingly undisturbed at $z=0$, but was stripped of stars at $\tlb \sim 6-8$ Gyr.}
  \label{fig:particleplots}
\end{figure*}

The left column of \Fig{particleplots} shows the distribution of stars around three example $\MBH(\Mstar)$-outlier galaxies that represent the range of environments in which our sample of 15 $\MBH(\Mstar)$-outliers reside at $z=0$ (from top-to-bottom, OutlierIDs 10, 8, and 6 in \Tab{outlier_table}). The top-left panel shows our most extreme (and lowest $\Mstar$) $\MBH(\Mstar)$-outlier with $\MBH \simeq 2 \times 10^8\Msun$ and $\Mstar \simeq 10^8 \Msun$, comprised of 85 stellar particles and one BH particle. This satellite is a mere 43 kpc from its much more massive ($\Mstar \simeq 6 \times 10^{11} \Msun$) host galaxy, but from the image there does not appear to be any sign of ongoing stellar stripping.\footnote{We have confirmed that there is indeed a stellar overdensity here -- the stellar mass density within 2 kpc is 10 times that between 2 and 10 kpc of this $\MBH(\Mstar)$-outlier.}  On the other hand, the middle-left panel shows another $\MBH(\Mstar)$-outlier that is clearly undergoing extensive stellar stripping due to tidal interactions. Finally, the $\MBH(\Mstar)$-outlier in the bottom-left panel is currently quite far (240 kpc) from its host and does not show any clear sign of ongoing tidal disturbances. Thus, these $\MBH(\Mstar)$-outliers can be found in many different dynamical states at $z=0$ (even though they are all indeed satellites).

In order to quantify their environments further, it is useful to determine the host of a given satellite galaxy. To this end, it is insufficient to simply select the central galaxy in its FoF group as the host. This is because in many cases a satellite may join the FoF group as a subhalo of a more massive satellite, and thus its dynamical history may be more closely linked to the more massive satellite than to the FoF central. To account for this situation, we compute for each satellite the tidal radius due to all of the more massive subhaloes in its FoF group using equation 7-84 of \citet{Binney1987} $-$ whichever subhalo yields the minimum tidal radius is then defined as the `host' galaxy. While this calculation is approximate (it assumes the haloes are point masses and the satellites are on circular orbits), it is sufficient to identify the true tidal perturber of a subhalo even though the computed tidal radius may be inaccurate.

In \Fig{environment} we show the distance of all 15 $\MBH(\Mstar)$-outliers to their respective host galaxies relative to the host virial radius ($D_{\rm host}/R_{\rm 200, host}$), along with the host virial mass, $M_{\rm 200, host}$, at $z=0$ (black solid lines).\footnote{We define the virial radius, $\R200$, as the radius within which the enclosed average density is 200 times the critical density of the universe at a given time; $\M200$ is the mass within $\R200$. Note that for subhaloes that are not the central of their FoF group, we approximate $\M200$ via the mean relation between $\M200$ and total (\textsc{subfind}) mass relation for centrals in the simulation.} Relative to the other satellites in the simulation that host BHs and have similar stellar mass ($\log_{10}(\Mstar/\Msun) \in [9.5,10.5]$; black dashed lines in \Fig{environment}), the $\MBH(\Mstar)$-outliers tend to be closer to their hosts (all within $0.5 R_{\rm 200,host}$; Kolmogorov-Smirnov (KS) test $p$-value, $p_{\rm KS}$, much less than 1 per cent between these two distributions) and to have hosts slightly (but not significantly; $p_{\rm KS} \approx 10$ per cent) more massive than average, indicating that $\MBH(\Mstar)$-outliers are typically subject to much stronger tidal forces than the majority of similar satellite galaxies. 


\subsection{Evolution}
\label{sec:evolution}

The right column of \Fig{particleplots} shows the evolutionary tracks of these three example $\MBH(\Mstar)$-outliers in the $\MBH-\Mstar$ plane obtained using the merger trees, with time running from light-yellow to dark-blue as indicated by the colour bar. For reference, the distribution for all galaxies at $z=0$ is shown underneath in grey (note that in the simulation this relation evolves towards higher $\Mstar$ from high redshift down to a lookback time $\tlb \sim 9$ Gyr ($z \sim 1.5$), after which it remains constant in time; see \Sec{others}). The subhalo is represented by a star or circle when it is a central or satellite, respectively. 

Our most extreme $\MBH(\Mstar)$-outlier (top-right panel of \Fig{particleplots}; OutlierID = 10) was indeed a much more massive galaxy in the past. Its stellar mass peaked at $3 \times 10^{10}\Msun$ at $z=2$ ($\tlb \simeq 10$ Gyr) before it became a satellite and gradually lost stellar mass until $z=0$. Indeed, since its $\Mstar$ at $z=0$ is very close to the resolution limit of the simulation, it is likely to be completely disrupted and lost within a Gyr after $z = 0$. In the middle-right panel, we see that the $\MBH(\Mstar)$-outlier with obvious ongoing stellar stripping (OutlierID = 8) only became a satellite at $\tlb\approx 1$ Gyr before quickly becoming an outlier at $z\approx0$ while it rapidly merges with its host galaxy (likely to be completely disrupted within the next several 100 Myr). Finally, the bottom-right panel shows that even the $\MBH(\Mstar)$-outlier satellite that looks relatively undisturbed at $z=0$ (OutlierID = 6) actually has lost most ($\approx 60$ per cent) of its stellar mass over the past 8 Gyr, but only after it became a satellite. For all three cases we have verified that this mass loss is due to the loss of stellar particles rather than stellar evolution. This evidence supports the idea that tidal stripping of more massive progenitor satellites may be the main cause of galaxies with overmassive BHs.

\begin{figure*}
    \centering
      \includegraphics[width=\textwidth]{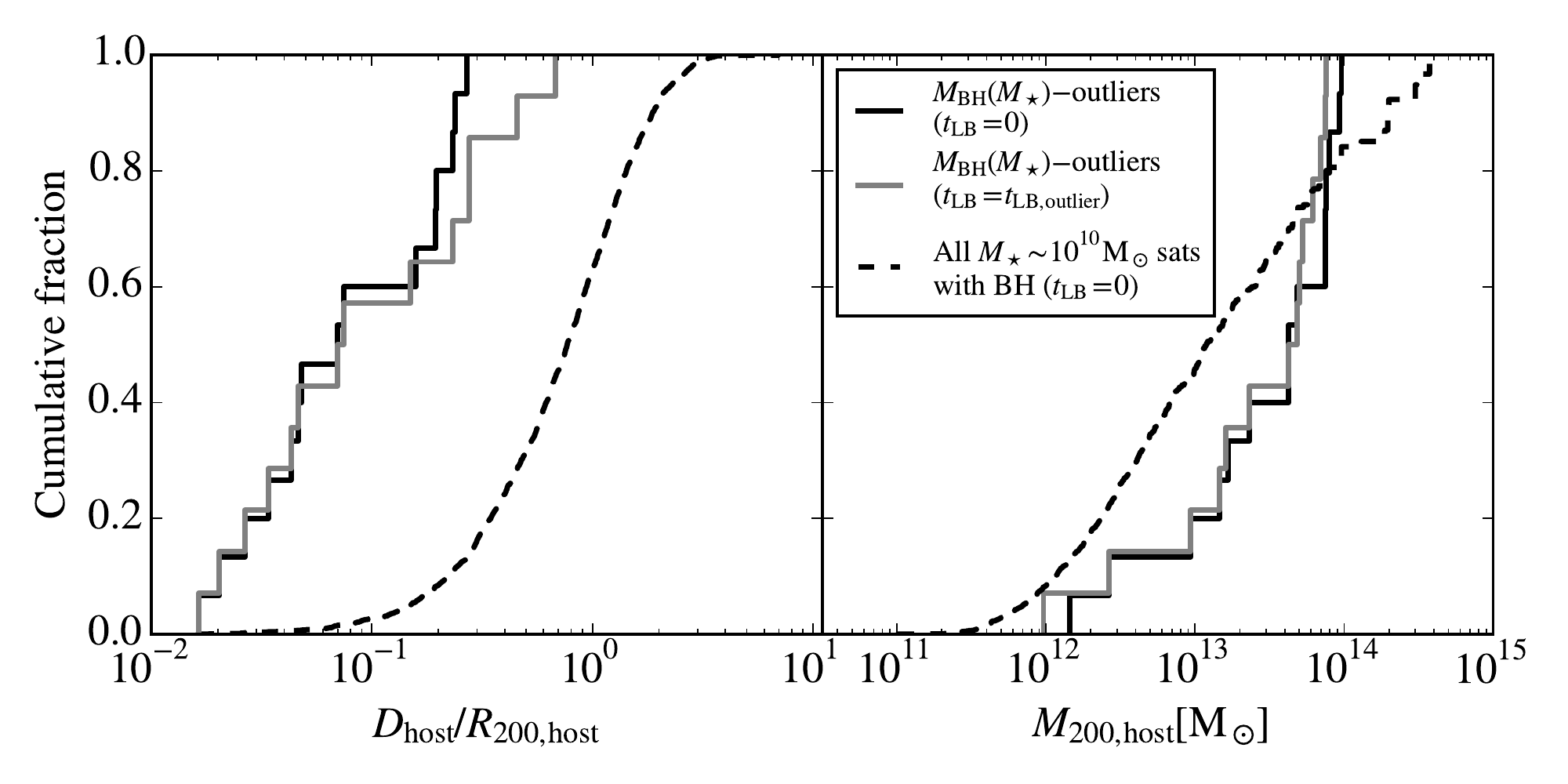}
  \caption{Cumulative distribution functions of the separation between satellite galaxies and their host galaxies normalized to $\R200$ of the host ({\it left panel}) and of $\M200$ of each satellite's host galaxy ({\it right panel}). Distributions are shown for $\MBH(\Mstar)$-outlier galaxies at $t_{\rm LB}=0$ (solid black lines) and when they became outliers in the instantaneous $\MBH-\Mstar$ relation ($t_{\rm LB, outlier}$; solid grey lines). Distributions at $t_{\rm LB}=0$ for all satellites with BHs and $\log_{10}(\Mstar/\Msun) \in [9.5,10.5]$ are shown as dashed lines. From the time at which they became $\MBH(\Mstar)$-outliers to $z=0$, $\MBH(\Mstar)$-outlier galaxies reside significantly closer to their host galaxies (all are within $0.5\R200$) than do typical satellite galaxies of similar stellar mass. }
  \label{fig:environment}
\end{figure*}

In \Fig{environment} we show the distance of all 15 $\MBH(\Mstar)$-outliers from their host galaxy relative to the virial radius of the host, as well as the host virial mass, measured at the time that each $\MBH(\Mstar)$-outlier became an outlier (i.e., when its $\MBH$ last rose higher than 1.5 dex above the (evolving) median $\MBH(\Mstar,z)$ relation; hereafter referred to as $t_{\rm LB, outlier}$; grey solid line). The distribution of separation from the host looks very similar to the $z=0$ case, albeit slightly more extended given some $\MBH(\Mstar)$-outliers may have only recently accreted on to the host at that time. However, all are within $\R200$ of the host, suggesting that tidal forces are likely to be responsible for the decrease in stellar mass since $t_{\rm LB, outlier}$. The $\M200$ distribution does not change significantly from $t_{\rm LB, outlier}$ to $z=0$. 

Note that there is one fewer $\MBH(\Mstar)$-outlier at $t_{\rm LB, outlier}$ than at $z=0$ in \Fig{environment}. This missing galaxy had $\log_{10}(\MBH/\MBHmed) > 1.5$ {\it before} it became a satellite, having gained in $\MBH$ and lost $\Mstar$ through stellar evolution at early times ($z=1-2$) and, as we shall see in the next section, has also not been stripped significantly of stars. With $\Mstar = 10^{10}\Msun$ and $\MBH \sim 6 \times 10^7 \Msun$, it is just at the edge of the general scatter in the $\MBH-\Mstar$ relation and thus may represent the high-mass tail end of `normal' $\MBH$ growth. Indeed, at $z=0$ it only just satisfies our $\MBH(\Mstar)$-outlier definition, lying $\approx 1.6$ dex above the median.


\subsection{Tidal stripping as the primary cause of anomalously high $\MBH(\Mstar)$}
\label{sec:tidalStripping}

\begin{figure}
  \centering
      \includegraphics[width=0.5\textwidth]{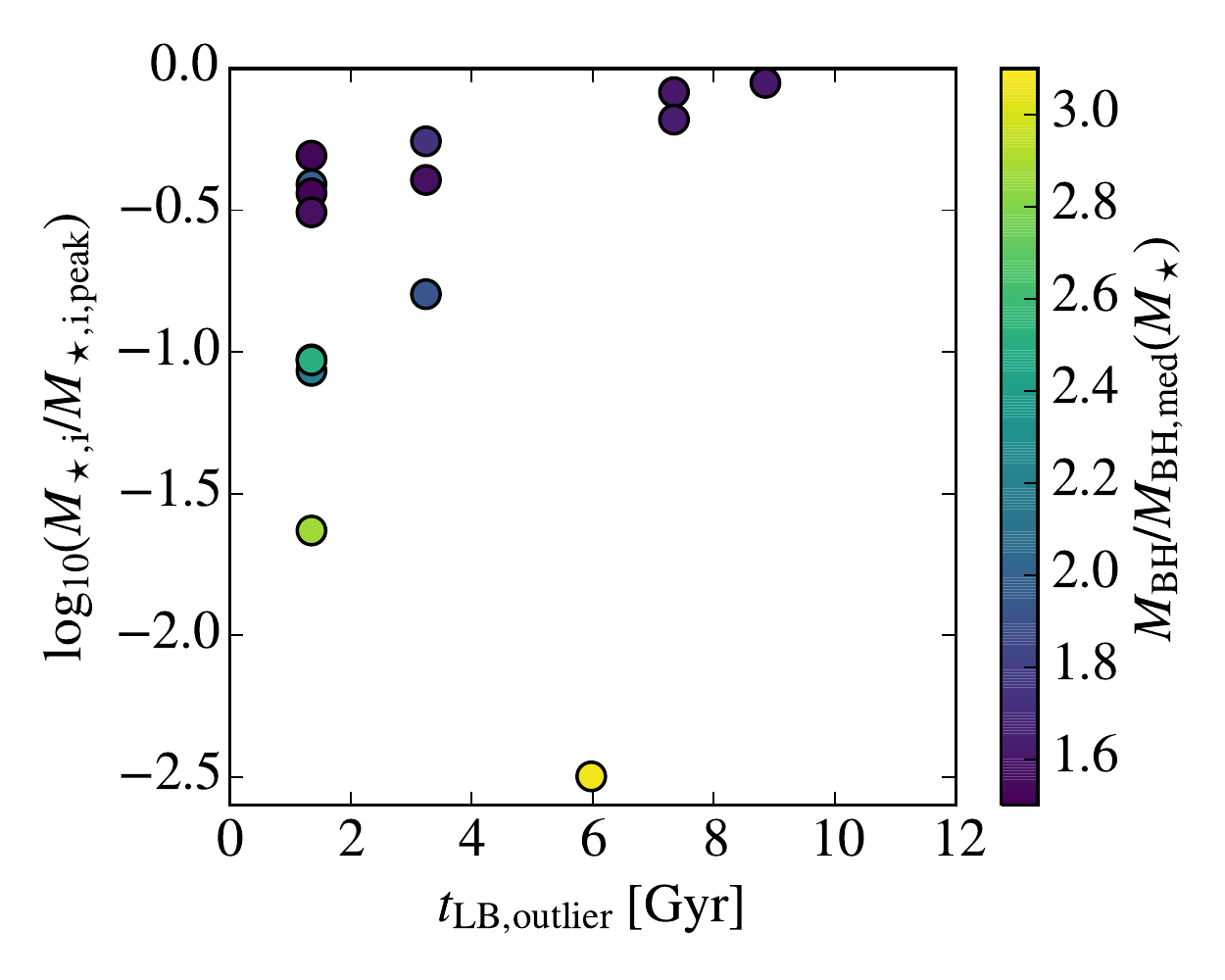}
  \caption{
Ratio of stellar mass over maximum stellar mass that a galaxy ever had as a function of elapsed time since it first became an outlier in the $\MBH-\Mstar$ relation, for all 15 $\MBH(\Mstar)$-outlier galaxies in EAGLE (see \Tab{outlier_table}). Points are coloured by $\MBH/\MBHmed$ at $z=0$. To remove the effect of mass loss due to stellar evolution, we use the sum of the {\it initial} masses of each galaxy's stellar particles, $\Mstari$. The most extreme $\MBH(\Mstar)$-outliers have been severely stripped, and became outliers within the past few Gyr.}
  \label{fig:lastTimeOnMedian_vs_stripping}
\end{figure}

If tidal stripping is responsible for creating $\MBH(\Mstar)$-outliers, then we can expect most of them to have become outliers recently, as those that begin to strip at earlier times are more likely to have been completely tidally disrupted by $z=0$. \Fig{lastTimeOnMedian_vs_stripping} shows the stellar mass loss (defined at the stellar mass at $z=0$ divided by the maximum stellar mass that it ever had, $\Mstari/\Mstaripeak$) as a function of the time since the $\MBH(\Mstar)$-outliers were last not outliers (i.e. the snapshot before $t_{\rm LB, outlier}$; see \Sec{evolution}). Note that here we use the sum of the {\it initial} stellar mass of each star particle, $\Mstari$, to ensure that the analysis is insensitive to any mass loss due to stellar evolution. 

Most of the EAGLE $\MBH(\Mstar)$-outliers became outliers in the past few Gyr and have been severely stripped by $z=0$, most having lost over 50 per cent of their maximum stellar mass. We do, however, find three galaxies that have been $\MBH(\Mstar)$-outliers for the past 7 to 9 Gyr, and have lost relatively little stellar mass. One of them was a $\MBH(\Mstar)$-outlier before becoming a satellite (as discussed in \Sec{evolution}), while the other two were already $>1.3$ dex above the median before becoming satellites, with a small amount of subsequent stellar mass loss pushing them just over the 1.5 dex cut soon thereafter. Indeed, the value of $t_{\rm LB, outlier}$ for these three galaxies is very sensitive to the definition of $\MBH(\Mstar)$-outliers as they are all only just above the 1.5 dex cut for most of their duration as satellites, rising only to $\sim 1.6$ dex at $z\sim 0$.

Near the bottom of \Fig{lastTimeOnMedian_vs_stripping} another $\MBH(\Mstar)$-outlier has lost nearly all of its $\Mstari$ by $z=0$ $-$ this is the most extreme $\MBH(\Mstar)$-outlier shown in the top row of \Fig{particleplots} (OutlierID = 10 in \Tab{outlier_table}). In \Fig{lastTimeOnMedian_vs_stripping} we see it has indeed lost 99.7 per cent of its peak (initial) stellar mass since it became a $\MBH(\Mstar)$-outlier 6 Gyr ago.

\begin{figure*}
  \centering
  \includegraphics[width=\textwidth]{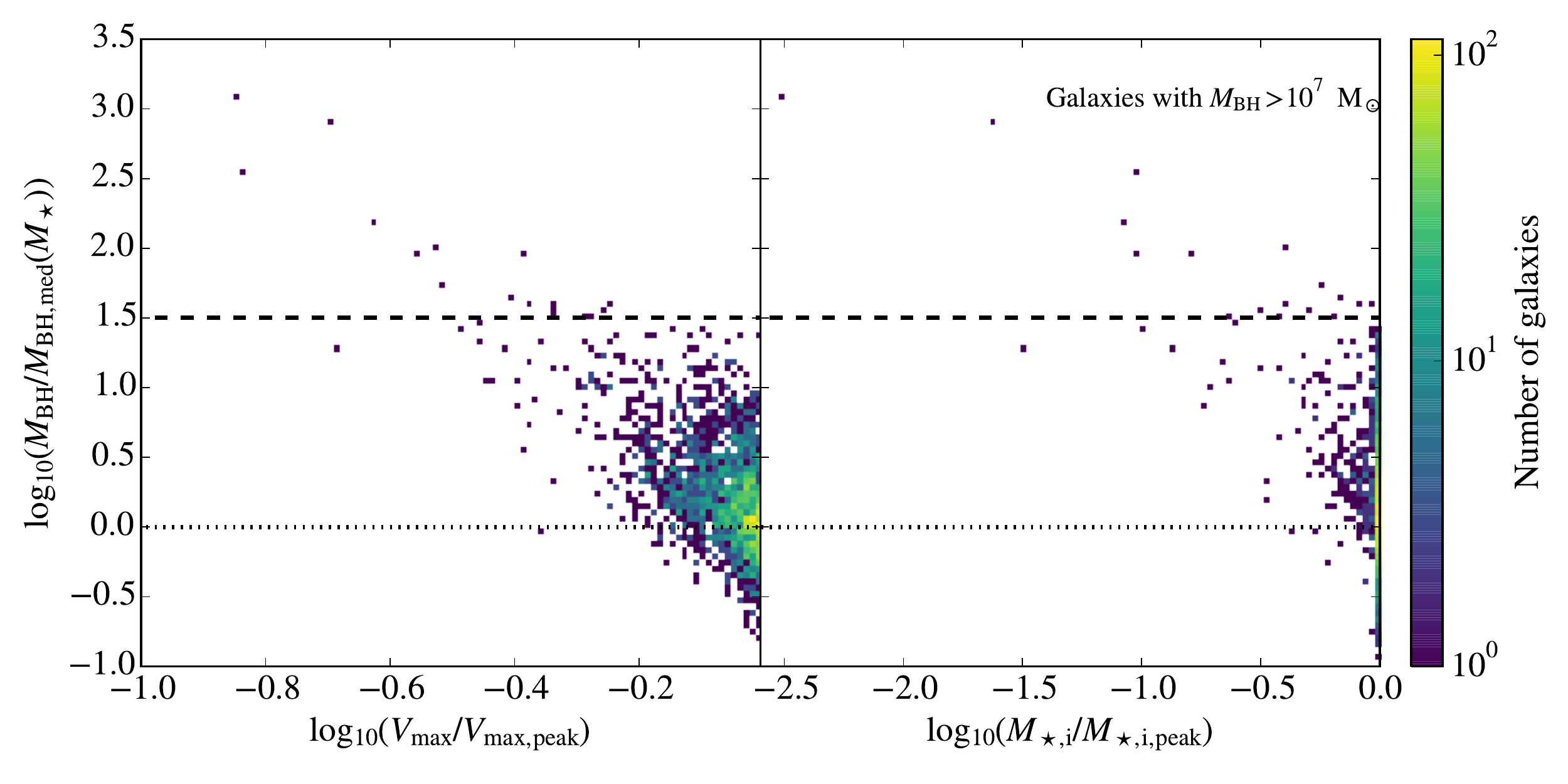}

  \caption{
Ratio of $\MBH$ over the median $\MBH$ for each galaxy's $\Mstar$ as a function of mass stripping proxy $\Vmax/\Vpmax$ (left-hand panel) and $\Mstari/\Mstaripeak$ (right-hand panel). Only galaxies with $\MBH > 10^7 \Msun$ at $z=0$ are shown in order to avoid BH seed mass resolution effects. Our 1.5 dex cut (definition of $\MBH(\Mstar)$-outliers) and the median $\MBH$ are shown as horizontal dashed and dotted lines, respectively. Galaxies that have significantly been stripped of stars tend to have $\MBH$ above the median value for their $\Mstar$, and indeed all $\MBH(\Mstar)$-outlier galaxies have lost some amount of stellar mass through tidal interactions.}

  \label{fig:resids_vs_stripping}
\end{figure*}

Another expectation of the tidal stripping hypothesis is that we should find a correlation between $\MBH/\MBHmed$ and stellar stripping. This test is especially important since it is insensitive to the $\MBH(\Mstar)$-outlier cut of $\log_{10}(\MBH/\MBHmed) > 1.5$. In \Fig{resids_vs_stripping} we plot this relation for all galaxies with $\MBH > 10^7 \Msun$, thus avoiding galaxies strongly affected by the finite BH seed mass (see \Fig{MBH_vs_Mstar}).

The left-hand panel of \Fig{resids_vs_stripping} shows $\MBH/\MBHmed$ as a function of the ratio between the maximum circular velocity, $\Vmax$, at $z=0$ and the highest value that it ever had, $\Vmaxpeak$, found by tracking its most massive progenitor back in time through the merger trees. There is a significant trend of stronger $\MBH(\Mstar)$-outliers with decreasing $\Vmax/\Vpmax$ for $\log_{10}(\Vmax/\Vpmax) < -0.2$ (Spearman rank-order correlation coefficient of $-0.6$ with $p$-value $\ll 1$ per cent), which is the regime where galaxies tend to begin losing stellar mass due to tidal stripping. 

In the right-hand panel of \Fig{resids_vs_stripping} we repeat the above analysis but use the ratio of the $z=0$ and peak initial stellar mass directly as a proxy for stellar stripping. Of the $\approx 2000$ galaxies with $\MBH > 10^{7}\Msun$, $24$ per cent have $\log_{10}(\Mstari/\Mstaripeak) < 0$, the vast majority of which are satellites.  All of our $\MBH(\Mstar)$-outlier galaxies have been stripped of stars, with the strongest outliers having lost the highest fraction of their maximum initial stellar mass. For galaxies with $\log_{10}(\Mstari/\Mstaripeak) < 0$, we obtain a Spearman rank-order correlation coefficient of -0.3 with $p \ll 1$ per cent, indicating a significant correlation between $\MBH/\MBHmed$ and stellar stripping.

The scatter in $\MBH/\MBHmed$ as a function of $\Vmax/\Vpmax$ is tighter than when plotted as a function of $\Mstar/\Mstaripeak$, which is surprising if stellar stripping is the direct cause of outliers in the $\MBH-\Mstar$ relation. This result is a consequence of the facts that $\MBH$ correlates more strongly with $\Vpmax$ than with $\Mstaripeak$ due to the strong dependence of $\MBH$ on halo binding energy \citep{Booth2010}, and that tidal stripping reduces $\Mstar$ and $\Vmax$ by roughly the same fraction for $\log_{10}(\Vmax/\Vpmax) < 0.8$ and $\Mstar > 10^8 \Msun$.

It is worth noting as well that these $\MBH(\Mstar)$-outlier (satellite) galaxies are also significant outliers in the relation between $\MBH$ and stellar velocity dispersion, $\sigma$, in the simulation. This is further evidence that they are inconsistent with being undisturbed relics of the high-$z$ universe, as such relic galaxies are expected to be outliers in $\MBH(\Mstar)$ but not in $\MBH(\sigma)$ \citep[][but see \Sec{others}]{Ferre-Mateu2015}.

Galaxies can also lose stellar mass through internal processes that cause the evaporation of stellar particles, such as three-body interactions with other stars or BHs (although the SPH softening prevents this so such processes are not captured in these simulations) or scattering off of large perturbations in the potential such as spiral arms or massive gas clumps. The impact that such processes have on galaxy stellar masses are not trivial to quantify. However, since these processes are internal to galaxies, one would expect both satellites and centrals to be affected equally. Thus, although it cannot be ruled out here, stellar evaporation is not expected to affect the relative offset between galaxies in the $\MBH-\Mstar$ relation. 

We thus conclude that tidal stripping is the dominant formation mechanism of galaxies with anomalously high BH masses in EAGLE.

\subsection{Early formation time as a secondary cause of anomalously high $\MBH(\Mstar)$}
\label{sec:others}

\begin{figure*}
  \centering

\subfloat{\includegraphics[width=.47\textwidth]{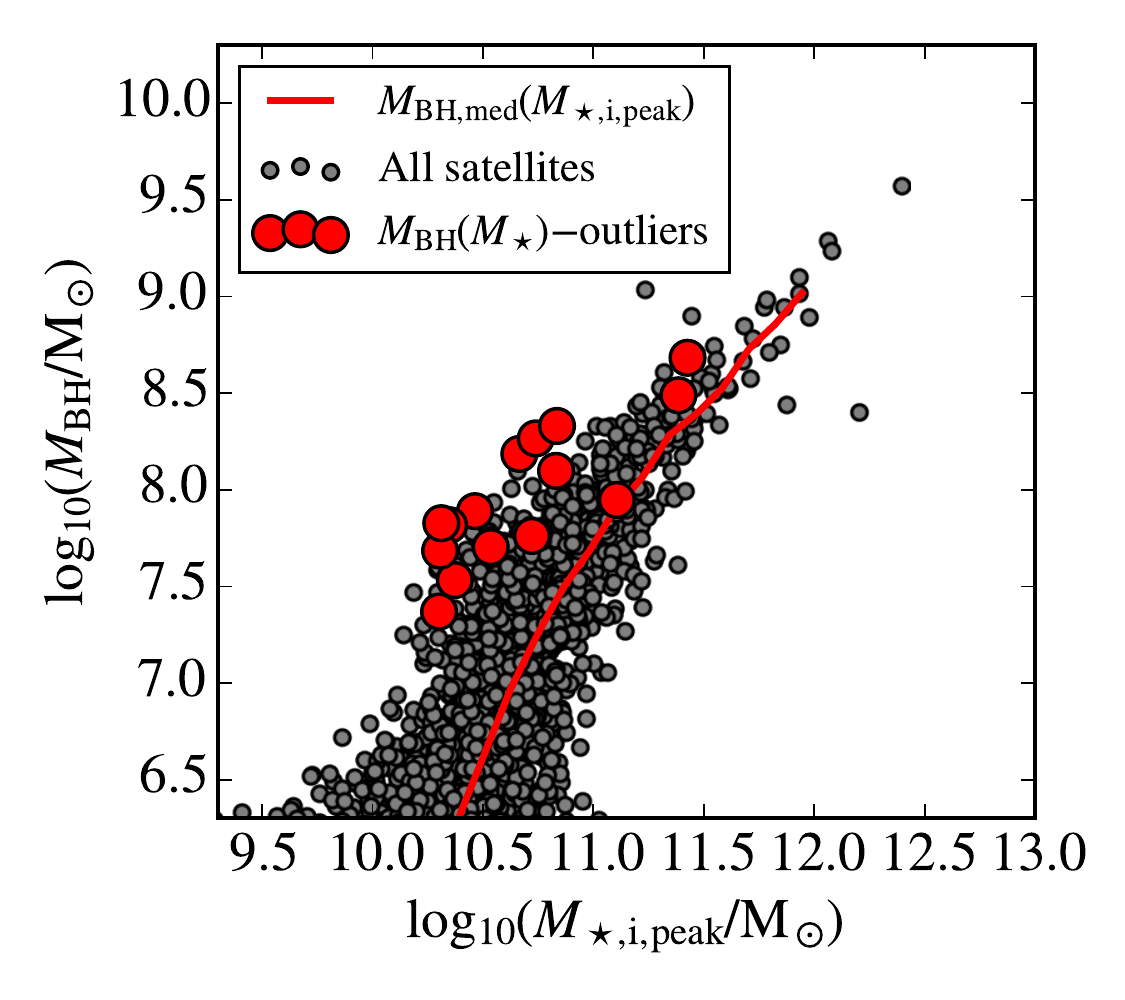}}
\subfloat{\includegraphics[width=.53\textwidth]{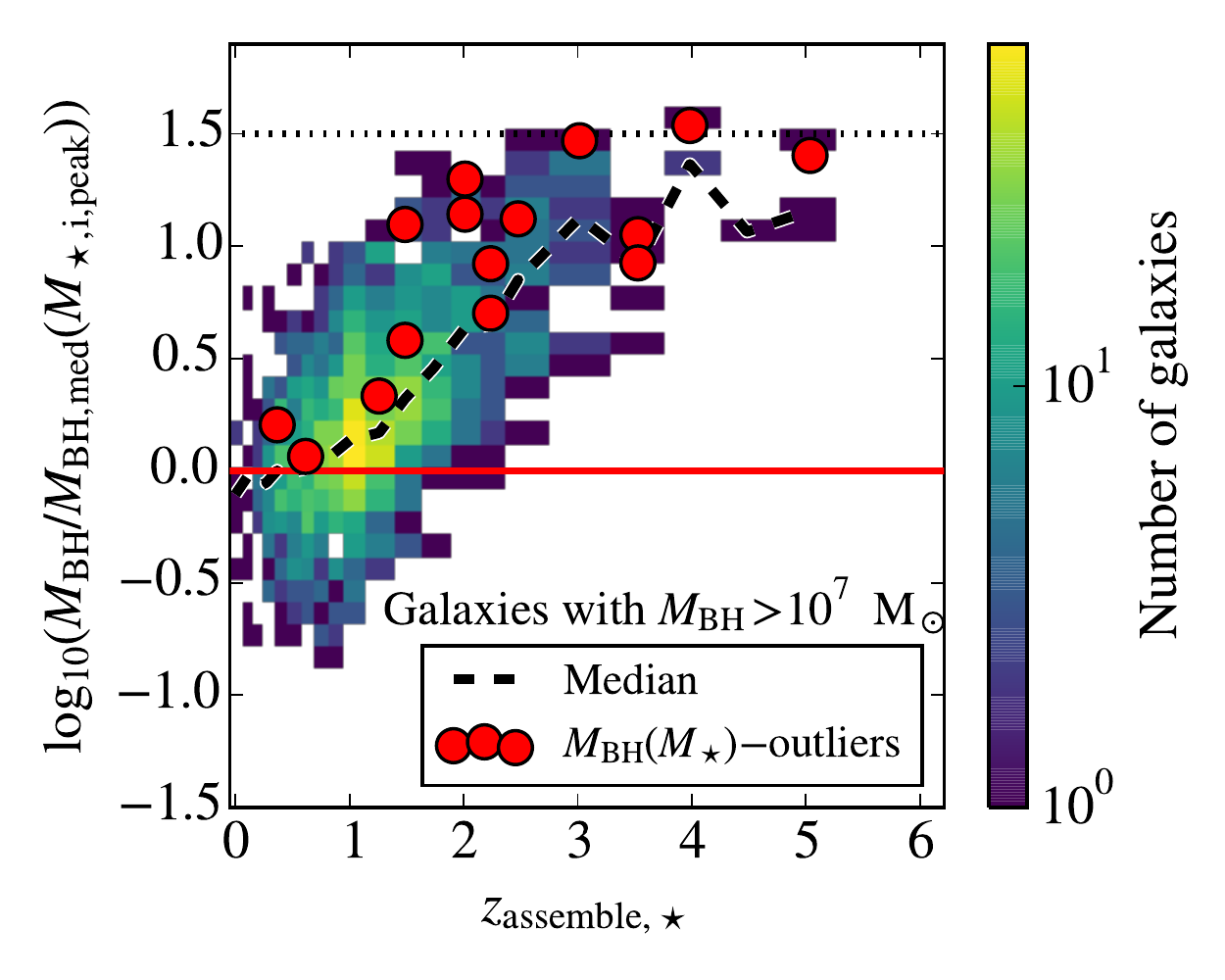}}

  \caption{{\it Left-hand panel:} $\MBH$ as a function of $\Mstaripeak$ for all satellite galaxies in EAGLE at $z=0$. The median $\MBH(\Mstaripeak)$ is shown as a red solid line while $\MBH(\Mstar)$-outlier galaxies are shown as red points. Even though the use of $\Mstaripeak$ removes the effect of stellar stripping, many $\MBH(\Mstar)$-outlier galaxies are $\approx 1$ dex above the median $\MBH(\Mstaripeak)$ relation. {\it Right-hand panel:} ratio of $\MBH$ over the median $\MBH(\Mstaripeak$) at $z=0$ as a function of stellar assembly redshift, $\zassstar$, defined as the earliest redshift at which $\Mstar(z) > 0.5 \Mstaripeak$, for galaxies with $\MBH>10^7 \Msun$. The dotted horizontal line denotes the $\MBH(\Mstar)$-outlier cut of 1.5 dex, while the solid red horizontal line denotes the median $\MBH(\Mstaripeak)$ at $z=0$. The black dashed line shows the median deviation above the median $\MBH(\Mstaripeak)$ relation (at $z=0$) as a function of $\zassstar$. After the effect of stellar stripping is removed, the remaining difference between the BH masses of the $\MBH(\Mstar)$-outliers and other galaxies of similar peak stellar mass is mostly due to their typically earlier stellar assembly times, as evidenced by the fact that the red points are close to the black dashed curve. }

  \label{fig:resids_vs_StarFormTime}
\end{figure*}

While all of our $\MBH(\Mstar)$-outliers [defined as $\log_{10}(\MBH/\MBHmed) > 1.5$] are tidally stripped satellites, tidal stripping may not be the only mechanism causing these galaxies to have unusually high $\MBH$. The left-hand panel of \Fig{resids_vs_StarFormTime} shows the $z=0$ relation between $\MBH$ and $\Mstaripeak$ for satellite galaxies, with the $\MBH(\Mstar)$-outlier galaxies highlighted in red. If tidal stripping were the only important mechanism in creating $\MBH(\Mstar)$-outliers, they would be expected to fall within the scatter in this relation. However, we find that most of them lie $\approx 1$ dex above the median $\MBH(\Mstaripeak)$ relation (red line), implying that, indeed, another physical mechanism must be affecting these galaxies.

An alternate explanation of $\MBH(\Mstar)$-outliers is that they are relics of the high-redshift Universe, when the $\MBH-\Mstar$ relation may have had a higher normalization. We test this scenario by measuring their stellar assembly redshifts, $\zassstar$, defined as the earliest redshift at which $\Mstari(z) \geq 0.5 \Mstaripeak$.

In the right-hand panel of \Fig{resids_vs_StarFormTime}, we plot the ratio between $\MBH$ and the median $\MBH(\Mstaripeak)$ relation at $z=0$ as a function of $\zassstar$. To avoid BH seed mass resolution effects we only consider galaxies with $\MBH > 10^7 \Msun$. Here we see a clear trend of increasing $\MBH/\MBHmedipeak$ with higher $\zassstar$, a trend that the $\MBH(\Mstar)$-outlier galaxies follow. Indeed, the vast majority of galaxies that formed before $z=1.5$ have $\log_{10}(\MBH/\MBHmed) > 0$. Of the 15 $\MBH(\Mstar)$-outlier galaxies, 10 (13) assembled most of their stars by $z=2$ $(1)$. Such early assembly times are atypical of satellite galaxies of similar stellar mass: the median $\zassstar$ of satellites with $\Mstaripeak = 10^{10-11}\Msun$ is $\approx 0.5$, much later than the $\MBH(\Mstar)$-outliers stellar assembly redshift of $\approx 2$.

\begin{figure*}
  \centering
  \includegraphics[width=0.8\textwidth]{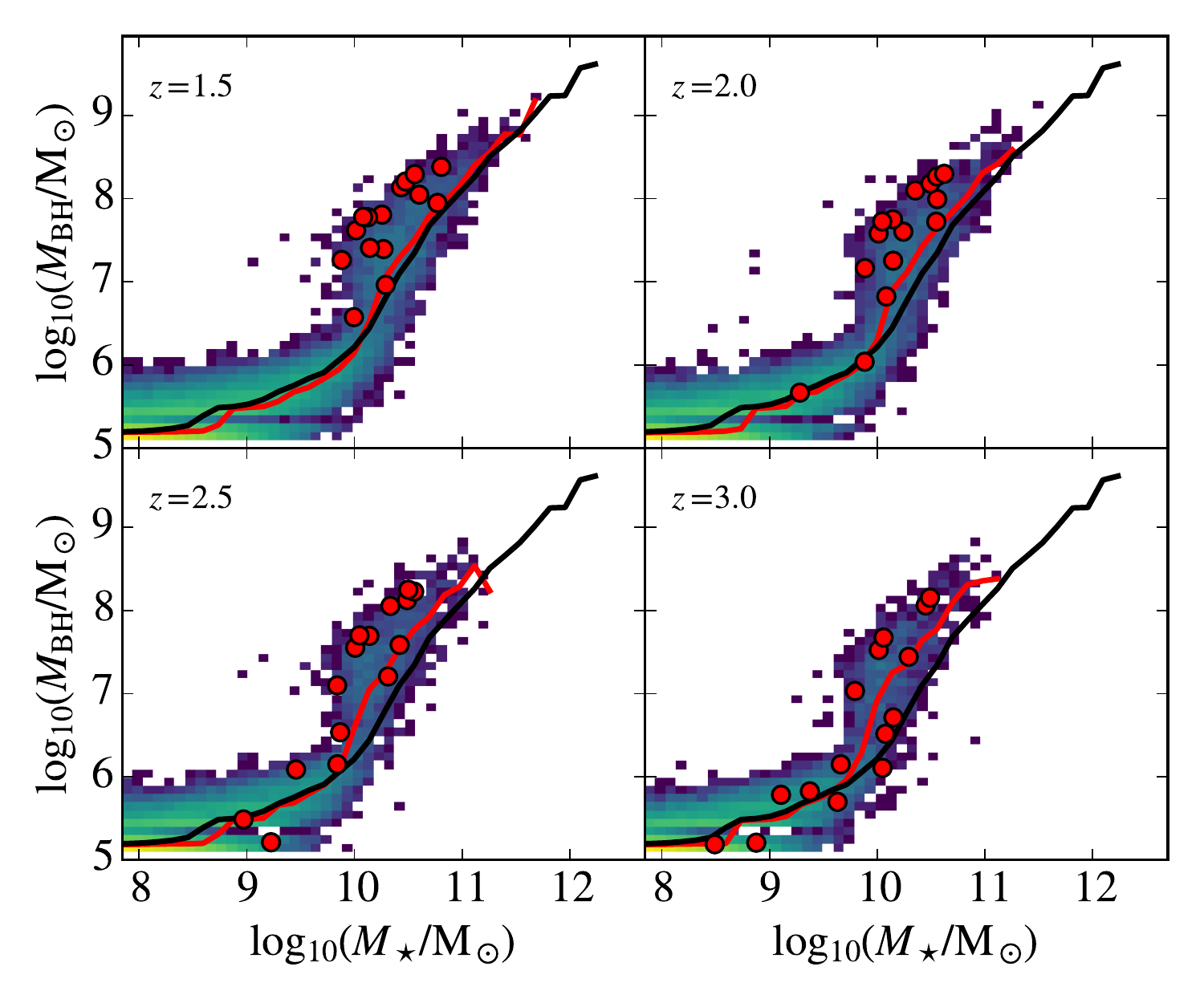}

  \caption{Evolution of the $\MBH-\Mstar$ relation with redshift (different panels). The main progenitors of the $z=0$ $\MBH(\Mstar)$-outliers are shown as red points. The median $\MBH(\Mstar)$ relation at each redshift is shown as a red curve. For reference, the median relation at $z=0$ is repeated in each panel as a black curve. The $\MBH-\Mstar$ relation evolved substantially from $z=3$ to $z=1.5$, dropping by nearly an order of magnitude at $\Mstar\sim10^{10}\Msun$. Progenitors of the $z=0$ $\MBH(\Mstar)$-outliers were not outliers from the $\MBH-\Mstar$ relation at $z=3$ and formed most of their stars and $\MBH$ from $z=3-2$ while the median $\MBH(\Mstar)$ relation was evolving, leaving them $0.5 - 1$ dex above the median relation by $z=1.5$, before most of them had become satellites. Galaxies that form most of their stars at high redshift tend to lie above the $\MBH-\Mstar$ relation at $z=0$ due to the evolution of the relation.}

  \label{fig:MBH_Mstar_evolution}
\end{figure*}

This trend with $\zassstar$ can be explained via the evolution of the median $\MBH(\Mstar)$ relation. \Fig{MBH_Mstar_evolution} shows $\MBH$ as a function of $\Mstar$ for all galaxies from $z=3-1.5$, the redshift range where this evolution is strongest. For $z>1.5$, the median $\MBH(\Mstar)$ was higher than it is at $z=0$ for $\Mstar\simgt 10^{10}\Msun$, with the largest deviation being approximately an order of magnitude at $z=3-4$ for $\Mstar \sim 10^{10}\Msun$. The main progenitors of the $\MBH(\Mstar)$-outliers (red points) are typical in terms of their BH and stellar mass at $z=3$, falling within the scatter of the relation at that redshift. Between $z=3$ and $z=2$, many of them grow rapidly in $\MBH$ while the median relation drops to lower BH masses, and already by $z=2$ most lie $\approx 0.5-1$ dex above the median relation. It was only after these processes occurred that these galaxies became satellites and began losing stellar mass, causing them to become the `extreme' $\MBH(\Mstar)$-outliers we find at $z=0$. For a more detailed discussion of the evolution of $\MBH$ in EAGLE, see \citet{Rosas-Guevara2016}.

It is interesting that we do not find more $\MBH(\Mstar)$-outliers with $\zassstar$ close to the more typical value of 0.5. One possibility is that the $\MBH(\Mstar)$-outliers that we {\it do} find are more resistant to complete tidal disruption than typical satellite galaxies. This would be the case if they are unusually compact, a hypothesis that is consistent with their early formation times \citep{Furlong2015_compact}, and one we explore in \Sec{compactness}. 

This early formation process occurred for central galaxies as well. While all of our $\MBH(\Mstar)$-outliers [defined as $\log_{10}(\MBH/\MBHmed) > 1.5$] are tidally stripped satellites, a slightly lower choice of $\MBH/\MBHmed$ threshold would have added central galaxies into our $\MBH(\Mstar)$-outlier sample, most of which have never been stripped. Indeed, a cut of 1.2 dex (thin dashed red line in \Fig{MBH_vs_Mstar}) yields 47 $\MBH(\Mstar)$-outlier galaxies, 10 of which are centrals (we will hereafter refer to these centrals as ``marginal $\MBH(\Mstar)$-outlier'' galaxies). The evolution of these marginal $\MBH(\Mstar)$-outliers in the $\MBH-\Mstar$ plane has been tracked in the same manner as the $\MBH(\Mstar)$-outliers in \Sec{evolution}. We find that none of them have been significantly stripped of stars $-$ most had a brief period of rapid BH growth at early times ($\tlb \simgt 12$ Gyr or $z \simgt 3$), and subsequently either grew following the same slope as the $z=0$ $\MBH-\Mstar$ relation, or simply stopped evolving. 

\begin{figure*}
  \centering
\includegraphics[width=\textwidth]{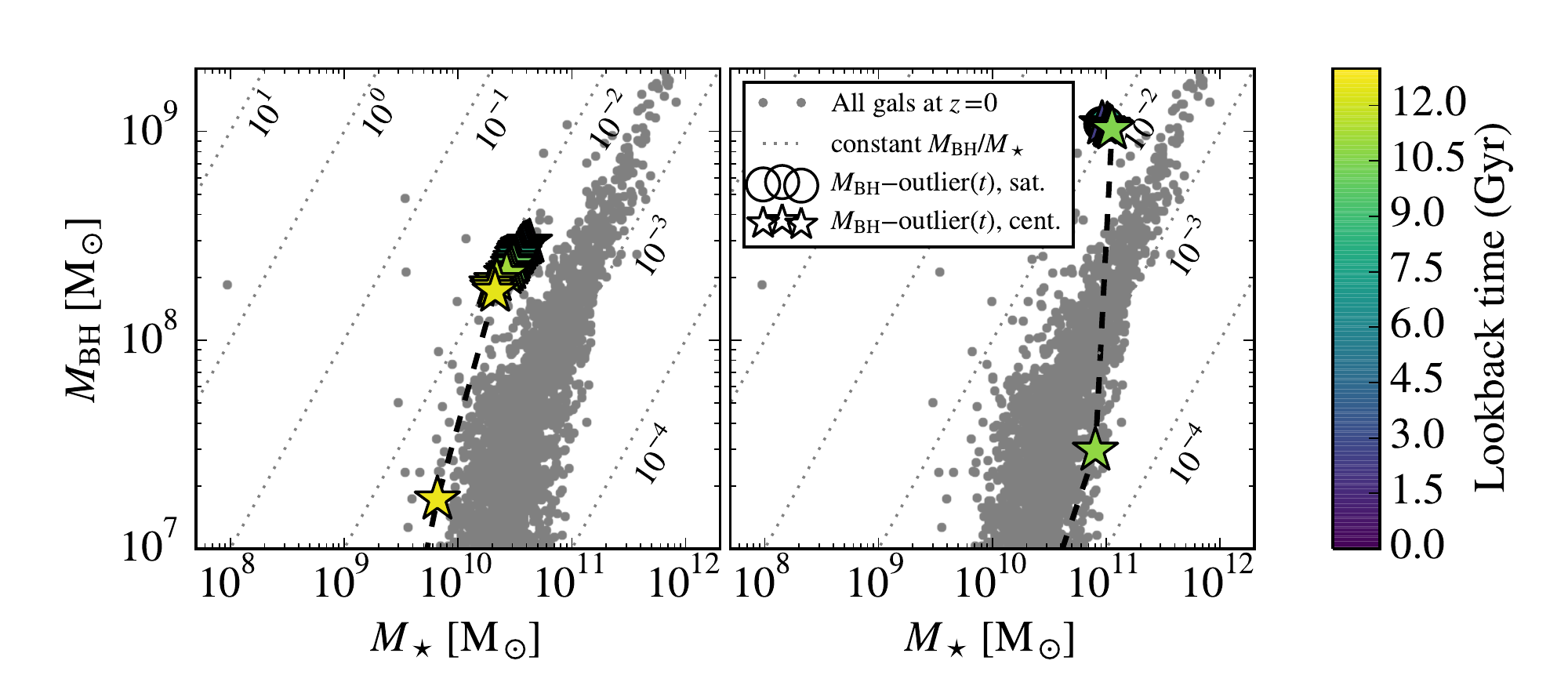}
  \caption{ As in the right column of \Fig{environment} but showing less extreme $\MBH(\Mstar)$-outliers that have not been stripped of stars. These are examples of high-$z$ `relic' galaxies. }
  \label{fig:others}
\end{figure*}

Two illustrative examples are shown in \Fig{others}. In the left-hand panel, we see one galaxy that already had a $\MBH/\Mstar$ ratio of 1 per cent by $\tlb = 12$ Gyr and thereafter evolved slowly with the same slope as the $z=0$ $\MBH-\Mstar$ relation. In the right-hand panel, we show a galaxy that grew very quickly in $\MBH$ up to a $\MBH/\Mstar$ ratio of 1 per cent at $\tlb = 11$ Gyr and remained stationary in the $\MBH-\Mstar$ plane thereafter. This result is consistent with the `high-$z$ relic' mechanism of $\MBH(\Mstar)$-outlier formation \citep{Ferre-Mateu2015}.

Indeed, the $\zassstar$ values of these 10 marginal $\MBH(\Mstar)$-outlier central galaxies are all higher than $z\simeq 2.5$ (see \Fig{resids_vs_StarFormTime}), making them bona fide relics of the early EAGLE universe. We note as well that these galaxies are not significant outliers from the median $\MBH(\sigma)$ relation (less than 0.5 dex above the median), consistent with the expected weak evolution of $\sigma$ since $z\approx2$ for individual galaxies \citep[e.g.][]{Cenarro2009, Hilz2012, Oser2012}. Thus, these relic galaxies were not strong outliers when they formed, but became (marginal) outliers from the $z=0$ $\MBH-\Mstar$ relation primarily due to the evolution of the $\MBH-\Mstar$ relation in time.

\subsection{ The relative importance of tidal stripping and early formation}
\label{sec:contributions}

\begin{figure*}
  \centering
  \includegraphics[width=\textwidth]{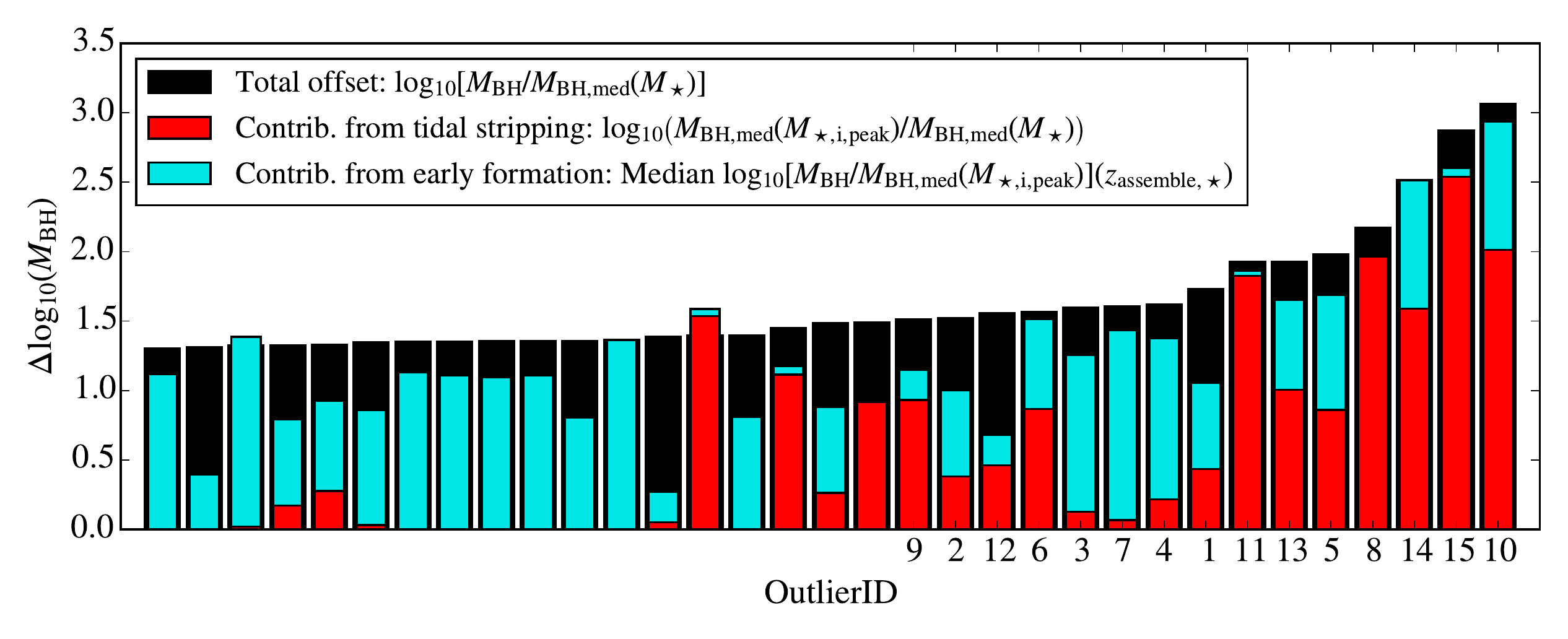}

  \caption{Individual contributions of tidal stripping (red bars) and early formation time (cyan bars, stacked) to the offset between $\MBH$ and $\MBHmed$ at $z=0$ (black bars) for galaxies with $\log_{10}(\MBH/\MBHmed) > 1.3$. The 15 most extreme $\MBH(\Mstar)$-outliers (with $\log_{10}(\MBH/\MBHmed) > 1.5$) are labelled with their OutlierIDs from \Tab{outlier_table}. Tidal stripping dominates over early formation for 10 of the 15 most extreme $\MBH(\Mstar)$-outliers, while stellar assembly time dominates for 80 per cent of galaxies with $\log_{10}(\MBH/\MBHmed) \in [1.3,1.5]$. Summed together, these two mechanisms account for on average 86 per cent of the offset for the most extreme outliers. }

  \label{fig:contributions}
\end{figure*}

We now quantify the individual contributions of stellar stripping and early formation times to the $\MBH$ offsets of $\MBH(\Mstar)$-outlier galaxies. We define the tidal stripping contribution by the difference between $\log_{10}(\MBHmed)$ and $\log_{10}(\MBHmedipeak)$ measured at the $\Mstar$ and $\Mstaripeak$ of each galaxy respectively. The contribution from early formation is defined as the median $\MBH/\MBHmedipeak$ at $z=0$ as a function of $\zassstar$ (the dashed line in the right-hand panel of \Fig{resids_vs_StarFormTime}), evaluated at the $\zassstar$ of each galaxy. These contributions are plotted in \Fig{contributions} for galaxies with $\log_{10}(\MBH/\MBHmed) > 1.3$, sorted by their $\MBH$ offset. Immediately we see that the contributions can vary wildly between galaxies, ranging from stripped late-assemblers to unstripped early-assemblers. Tidal stripping dominates for 10 of the 15 $\MBH(\Mstar)$-outlier galaxies (including the four most extreme outliers; $\log_{10}[\MBH/\MBHmed] > 2$), while early formation is more important for $\approx 80$ per cent of the 18 less extreme $\MBH(\Mstar)$-outliers (with $\log_{10}(\MBH/\MBHmed) \in [1.3,1.5]$)\footnote{The same result is found for the 32 galaxies with $\log_{10}(\MBH/\MBHmed) \in [1.2,1.5]$.}. The contribution from stripped galaxies found here may even be an underestimate, since with increased mass resolution it would be possible to track satellites down to smaller masses and thus for longer times during the stripping process before being lost by the subhalo finder, increasing the number of stripped galaxies at any given time.

Summed together, the contributions from tidal stripping and early stellar assembly times account for on average 86 per cent of the offset in $\MBH(\Mstar)$ at $z=0$ for the extreme $\MBH(\Mstar)$-outliers, leaving them on average $0.3 \pm 0.2$ dex above the median $\MBH(\Mstaripeak)$ relation after these two effects have been taken into account, well within the 95 percentile scatter of $\approx 0.6$ dex around the median at $\Mstaripeak \sim 10^{10.5} \Msun$. We attribute this additional offset of $0.3 \pm 0.2$ dex above the median to a selection bias. At fixed $\Mstaripeak$, the deficit in $\Mstar$ at $z=0$ required to be a $\MBH(\Mstar)$-outlier by our definition increases strongly with decreasing $\MBH$. Thus, we expect the $\MBH(\Mstar)$-outliers to be biased to high $\MBH$ even with the effects of tidal stripping and early formation taken into account.

These results are qualitatively consistent with the properties of galaxies with overmassive BHs in the (142 Mpc)$^3$ cosmological simulation HorizonAGN, where the most extreme outliers in the simulated $\MBH-\Mstar$ relation are stripped satellites of more massive host galaxies while relic galaxies have the highest BH masses of the central galaxies but do not have as extreme BH masses as stripped satellites \citep{Volonteri2016}.

We conclude that, while the extreme $\MBH(\Mstar)$-outliers result primarily from tidally-induced stellar stripping, it is certainly possible for galaxies to simply form a high-mass BH at very early times and then stop evolving, becoming `relics' of the early Universe by $z=0$. Indeed, the most extreme outliers formed relatively early and already had relatively massive BHs prior to losing stars due to stellar stripping.  

\section{Relation to compact galaxies}
\label{sec:compactness}

The formation of UCDs has recently received much attention. The general picture that is emerging posits that the highest mass UCDs ($\Mstar \simeq 10^8 \Msun$) may be the surviving nuclei of dwarf elliptical galaxies whose outer parts have been tidally `threshed' by more massive host galaxies \citep[e.g.,][]{Bekki2003,Brodie2011, Chilingarian2011, DaRocha2011,Pfeffer2014,Norris2014,Norris2015}. Recently, \citet{Mieske2013} found that massive UCDs ($\Mstar > 10^7 \Msun$) have excess dark mass which can be explained by massive BHs (or a bottom-heavy IMF), which would offset them by up to 2 dex above the $\MBH-\sigma$ and $\MBH-L_{\rm bulge}$ relations. Thus, if the existence of their central BHs is confirmed, they would be strong $\MBH(\Mstar)$-outliers. Indeed, the confirmation of a central BH in M60-UCD1 \citep[][see also \Fig{MBH_vs_Mstar}]{Seth2014} suggests that the tidal formation mechanism may dominate UCD formation, at least for the most massive UCDs.

As our $\MBH(\Mstar)$-outliers are the tidally threshed nuclei of more massive progenitors, it is thus natural to ask if there is some connection between the $\MBH(\Mstar)$-outliers and compact galaxies in the simulation, and whether this can be used to shed some light on the formation mechanism of UCDs. Note that at our resolution we cannot resolve galaxies as low in mass ($\Mstar$ typically between $2 \times 10^6$ and $10^8 \Msun$) or as small (effective radii $ \simlt 100$ pc) as UCDs. Indeed, we find only one $\MBH(\Mstar)$-outlier with $\Mstar \sim 10^8 \Msun$, and even this one has $\MBH$ an order of magnitude higher than what would be consistent with UCDs (OutlierID=10 in \Tab{outlier_table}). However, if UCDs are the tidally threshed nuclei of more massive progenitors, then we can likely catch these progenitors after they have started to lose mass, but before they are stripped to masses below our resolution limit. Indeed, if the $\MBH(\Mstar)$-outliers tend to be more compact than average, then they could very well be on their way to becoming compact galaxies similar to UCDs.

To this end, we now investigate whether the $\MBH(\Mstar)$-outlier galaxies are unusually compact given their $\Mstar$, and if so, whether compact galaxies at $z=0$ generally have been stripped of stars. The evolution of galaxy sizes in EAGLE, including high-$z$ compact galaxies, has already been discussed by \citet{Furlong2015_compact}; thus in this section we perform a qualitative treatment of compact galaxies, looking only for general trends. 

We define `compactness' as the ratio between the 3D stellar half-mass radius of a galaxy, $\Rhalf$, and the median value for galaxies with the same $\Mstar$, $\Rhalfmed$; galaxies with lower ratios are more compact. Only galaxies with $\Mstar > 10^{9} \Msun$ are considered here since below this limit the measured size is unconverged in the Ref-L0100N1504 simulation for typical galaxies (S15). For completeness we also include the most extreme $\MBH(\Mstar)$-outlier (OutlierID = 10, $\Mstar \simeq 10^8 \Msun$); its inclusion does not affect our results. Our $\MBH(\Mstar)$-outlier galaxies have $\Rhalf \sim 1 - 3$ kpc, which approaches the softening length of 0.7 proper kpc. Thus, their sizes may still be affected by resolution and should be taken as upper limits. 

\begin{figure}
  \centering
      \includegraphics[width=0.5\textwidth]{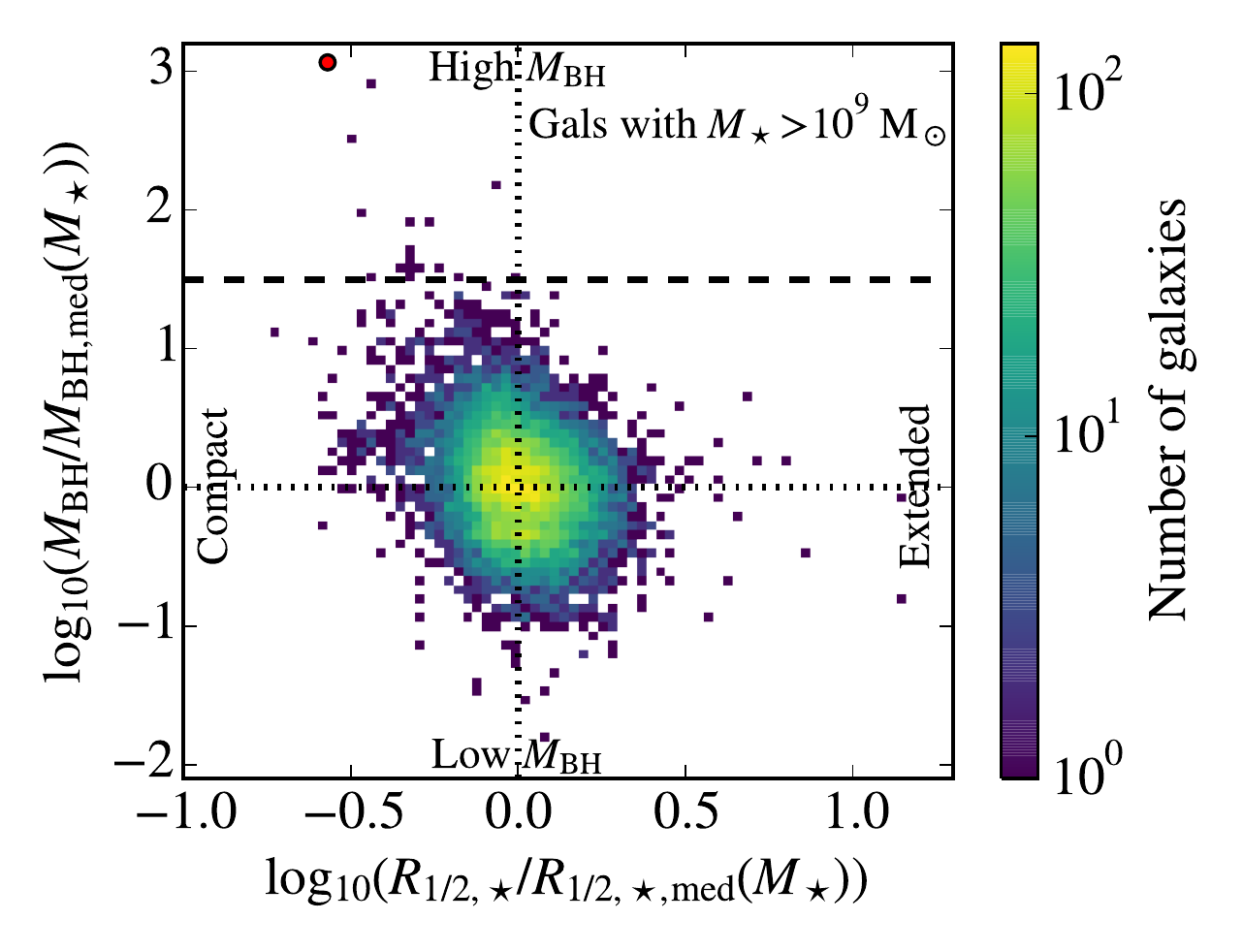}
  \caption{2D histogram of the number of dex above the median $\MBH(\Mstar)$ relation as a function of compactness for galaxies with $\Mstar > 10^9 \Msun$. Our $\log_{10}(\MBH/\MBHmed) > 1.5$ cut defining $\MBH(\Mstar)$-outliers is shown as a dashed horizontal line, while the median $\MBH$ and $\Rhalf$ are shown as horizontal and vertical dotted lines, respectively. For completeness we include the most extreme $\MBH(\Mstar)$-outlier (OutlierID = 10) as a red circle. More compact galaxies tend to have higher-than-average $\MBH$, especially those with half-mass radii more than 0.3 dex below the median value for their stellar mass.}
  \label{fig:compactness_vs_resids}
\end{figure}

\Fig{compactness_vs_resids} shows the relation between $\MBH/\MBHmed$ and $\Rhalf/\Rhalfmed$. We find that the $\MBH(\Mstar)$-outlier galaxies are significantly more compact than the median size given their stellar masses, with most of them $\sim 0.2-0.5$ dex below the median. In general, we see that galaxies more compact than $\log_{10}(\Rhalf/\Rhalfmed) = -0.3$ tend to have higher $\MBH$ than expected given their $\Mstar$ (with a median $\log_{10}(\MBH/\MBHmed) \approx 0.35$). On the other hand, galaxies with $\log_{10}(\MBH/\MBHmed) > 1$ tend to be more compact than the median (with median $\log_{10}(\Rhalf/\Rhalfmed) \approx -0.18$). Thus, there does seem to be a strong connection between overmassive BHs and compactness. This result supports the hypothesis that UCDs are the tidally threshed nuclei of more massive progenitor galaxies. Note, however, that this result does not imply that all compact galaxies have overmassive BHs. To answer this question one would need to resolve $\Rhalf \simlt 1$ kpc (corresponding to $\log_{10}(\Rhalf/\Rhalfmed) < -0.6$  at $\Mstar\sim 10^{10}\Msun$), which is not possible at this resolution.

\begin{figure}
  \centering

\includegraphics[width=.5\textwidth]{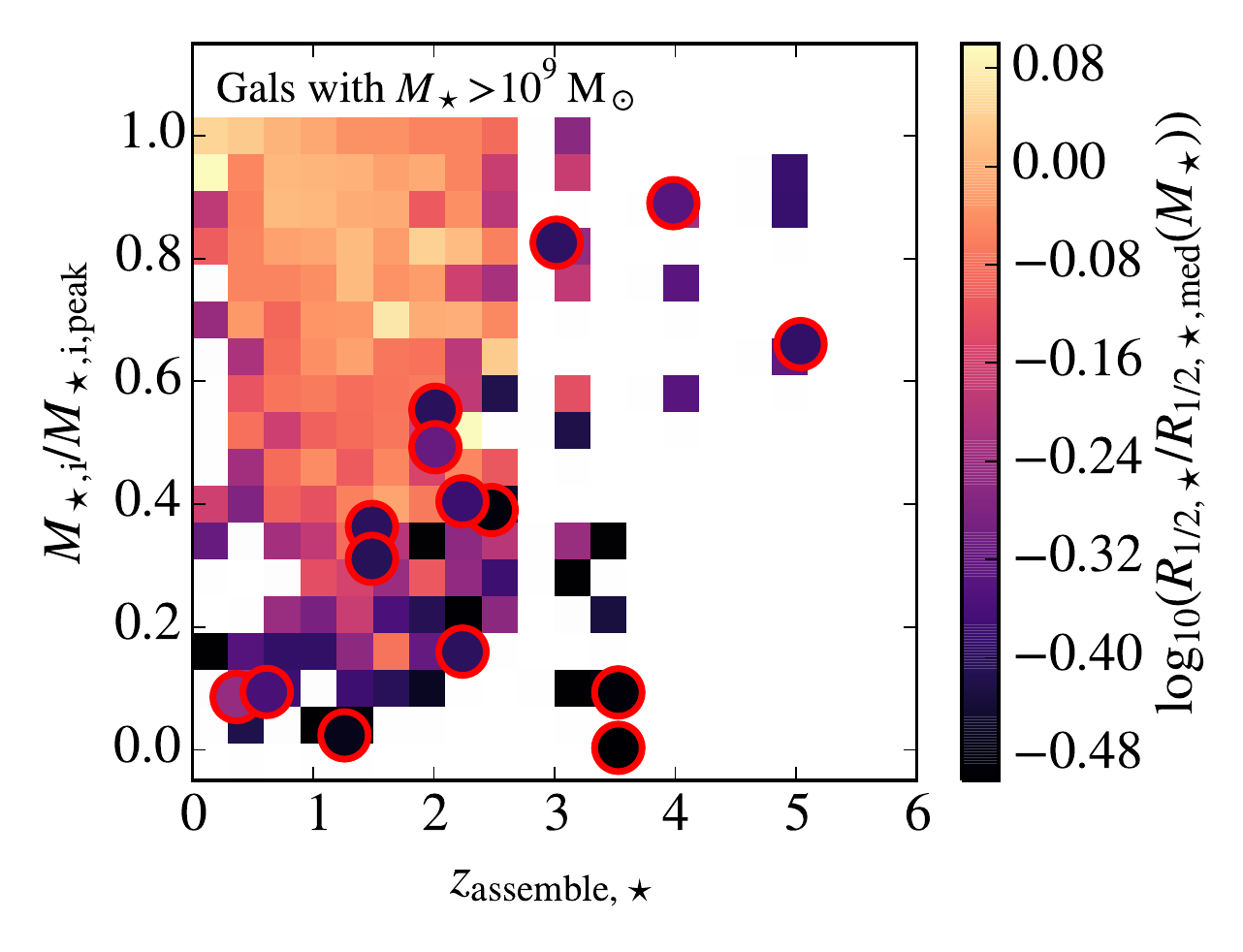}

  \caption{ Median compactness in bins of $\Mstari/\Mstaripeak$ and $\zassstar$ for all galaxies with $\Mstar > 10^9 \Msun$. The $\MBH(\Mstar)$-outlier galaxies are shown as red circles. Galaxy compactness increases with a higher degree of stellar stripping and with an earlier assembly redshift. The $\MBH(\Mstar)$-outlier galaxies are compact because they assembled their stars at high-$z$ and/or were tidally stripped of stars after becoming satellite galaxies. }

  \label{fig:compactness_vs_stripping}
\end{figure}

Since the combination of a high stellar assembly redshift and, most importantly, tidal stripping are responsible for creating $\MBH(\Mstar)$-outlier galaxies, we now investigate if the same mechanisms cause galaxies to be more compact in general.  \Fig{compactness_vs_stripping} shows $\Rhalf/\Rhalfmed$ as a function of $\Mstari/\Mstaripeak$ and $\zassstar$, for all galaxies with $\Mstar > 10^9 \Msun$, regardless of whether or not they harbour a BH. We see that, indeed, galaxies that have lost more than 40 per cent of their stellar mass and/or have assembled more than half of their stellar mass before $z=2.5$ tend to be more compact than expected given their (current) mass, with median $\Rhalf/\Rhalfmed \approx -0.2$. For reference, we overplot the $\MBH(\Mstar)$-outlier galaxies, finding that 13 of 15 lie in at least one of these two regimes. Thus, the compactness of the $\MBH(\Mstar)$-outlier galaxies, as is the case for their high BH masses, is caused by a combination of their early assembly times and stellar stripping.

The fact that all of our (stripped) $\MBH(\Mstar)$-outlier satellites are compact has at least two possible explanations: (1) a physical explanation where more extended satellite galaxies are not able to withstand significant tidal stripping before being completely disrupted, or (2) a numerical explanation where more extended $\MBH(\Mstar)$-outliers {\it do} exist at $z=0$, but are undetectable by our halo finder (and in observations) because they represent much weaker overdensities relative to the host galaxy. More sophisticated subhalo finding techniques would be required to detect such extended satellite galaxies, a task beyond the scope of this work. 

We conclude that galaxies with overly massive BHs tend to be more compact than typical galaxies of the same stellar mass, as expected for the tidal threshing hypothesis of UCD formation as well as in the early formation scenario. $\MBH(\Mstar)$-outlier galaxies are compact due to a combination of their high degree of stellar stripping and their early formation times. While this tidal threshing mechanism clearly occurs within the framework of EAGLE, higher-resolution simulations would be needed to follow UCD formation in detail.

\section{Summary and Conclusions}
\label{sec:conclusions}

We have investigated the existence and evolution of galaxies that are positive outliers relative to the $z=0$ $\MBH-\Mstar$ relation using the (100 Mpc)$^3$ EAGLE cosmological hydrodynamical simulation. Our main conclusions are as follows.

\begin{itemize}
	\item Positive outliers from the $\MBH-\Mstar$ relation (referred to as $\MBH(\Mstar)$-outliers) similar to those presented in the recent observational literature exist in EAGLE. Most $\MBH(\Mstar)$-outliers have stellar masses $\Mstar \sim 10^{10} \Msun$ and black hole masses $\MBH \sim 10^8 \Msun$. These mass ranges are similar to observed $\MBH(\Mstar)$-outliers NGC 4486B, NGC 4342 and S536. However, we cannot make predictions for the existence of $\MBH(\Mstar)$-outliers as massive as, e.g., NGC 1277 due to the limited box size, or as low in mass as ultracompact dwarf (UCD) galaxies due to the finite resolution of EAGLE (\Fig{MBH_vs_Mstar}).
	\item The 15 most extreme $\MBH(\Mstar)$-outliers (defined as those with $\log_{10}[\MBH/\MBHmed] > 1.5$) are satellites of larger host galaxies, each residing within half the virial radius of its host halo (Fig. \ref{fig:environment}).
	\item These extreme $\MBH(\Mstar)$-outliers became outliers through a combination of early stellar assembly times (Figs. \ref{fig:resids_vs_StarFormTime} and \ref{fig:MBH_Mstar_evolution}) and subsequent extensive stellar stripping due to tidal forces from their host haloes (Figs. \ref{fig:particleplots} and \ref{fig:resids_vs_stripping}). Tidal stripping is the dominant mechanism responsible for the overmassive BHs in 67 per cent of the extreme outliers (Fig. \ref{fig:contributions}), a fraction that may increase for higher-resolution simulations. Some $\MBH(\Mstar)$-outliers have only recently begun to undergo stripping while others are survivors of slow tidal stripping occurring over several Gyr (Figs. \ref{fig:particleplots} and \ref{fig:lastTimeOnMedian_vs_stripping}). The most extreme $\MBH(\Mstar)$-outliers are currently undergoing severe stellar disruption. 
        \item Early formation is more important than tidal stripping for causing the $\MBH$ offsets in 80 per cent of the 32 less extreme $\MBH(\Mstar)$-outliers ($1.2 < \MBH/\MBHmed < 1.5$; \Fig{contributions}). Such galaxies formed with overmassive BHs at high redshift $(z > 2$) when the normalization of the $\MBH-\Mstar$ relation was higher (Figs. \ref{fig:resids_vs_StarFormTime} and \ref{fig:MBH_Mstar_evolution}). Of these early-forming galaxies, 10 are centrals that subsequently either evolved parallel to the $z=0$ $\MBH-\Mstar$ relation or remained unchanged until $z=0$, becoming `relics' of the high-$z$ EAGLE universe (Fig \ref{fig:others}).
        \item Together, the combination of tidal stripping and early stellar assembly times accounts for an average of 86 per cent of the offset above the median $\MBH(\Mstar)$ for the extreme $\MBH(\Mstar)$-outliers (\Fig{contributions}). 
	\item The extreme $\MBH(\Mstar)$-outliers are amongst the most compact galaxies in the simulation, with stellar half-mass radii, $\Rhalf$, typically $0.2-0.5$ dex smaller than the median value for other galaxies of similar stellar mass, $\Rhalfmed$, making them ideal candidates for UCD progenitors. Similarly, galaxies with $\Mstar > 10^9 \Msun$ that are more than 0.3 dex below $\Rhalfmed$ tend to host overmassive BHs. These $\MBH(\Mstar)$-outliers, and galaxies with $\Mstar > 10^9\Msun$ in general, become compact via a combination of early formation and/or tidal stripping (Figs. \ref{fig:compactness_vs_resids} and \ref{fig:compactness_vs_stripping}).

\end{itemize}

\section*{Acknowledgements}

The authors are grateful to the anonymous referee for a very constructive report that improved the quality of the paper. CB thanks Remco van den Bosch, Tiago Costa, and Joseph Silk for helpful comments and suggestions, and Stuart McAlpine for many useful discussions regarding spurious galaxies in EAGLE. This work used the DiRAC Data Centric system at Durham University, operated by the Institute for Computational Cosmology on behalf of the STFC DiRAC HPC Facility (www.dirac.ac.uk). This equipment was funded by BIS National E-infrastructure capital grant ST/K00042X/1, STFC capital grants ST/H008519/1 and ST/K00087X/1, STFC DiRAC Operations grant ST/K003267/1 and Durham University. DiRAC is part of the National E-Infrastructure. RAC is a Royal Society University Research Fellowship. We also gratefully acknowledge PRACE for awarding us access to the resource Curie based in France at Tr$\grave{\rm e}$s Grand Centre de Calcul. This work was sponsored by the Dutch National Computing Facilities Foundation (NCF) for the use of supercomputer facilities, with financial support from the Netherlands Organization for Scientific Research (NWO). The research was supported in part by the European Research Council under the European Union's Seventh Framework Programme (FP7/2007-2013)/ERC grant agreement 278594-GasAroundGalaxies and 267291-Cosmiway. This research was funded by the Interuniversity Attraction Poles Programme initiated by the Belgian Science Policy Office (AP P7/08 CHARM). This research made use of {\sc astropy}, a community-developed core {\sc python} package for Astronomy \citep{Astropy2013}.




\bibliographystyle{mnras} 
\bibliography{bhstrip} 


\bsp	
\label{lastpage}
\end{document}